\documentclass[11pt, oneside]{article}   	
\usepackage{geometry}
\usepackage{amsmath}
\numberwithin{equation}{section}

\usepackage{graphicx}
\usepackage{subfigure}
\usepackage{amssymb}
\usepackage{mathrsfs}  
\usepackage[shortlabels]{enumitem}
\usepackage{mathtools}
\usepackage{caption}
\usepackage{cite}
\usepackage{hyperref}
	\hypersetup{
	    colorlinks=true,
	    anchorcolor=black,
	    urlcolor=black,
	    citecolor=red,
	    linkcolor=red}

\def\be{\begin{equation}}
\def\ee{\end{equation}}
\def\bsub{\begin{subequations}}
\def\esub{\end{subequations}}
	    
\def\m{\mu}
\def\n{\nu}
\def\a{\alpha}
\def\b{\beta}

\def\la{\lambda}

\def\ka{\varkappa}

\def\ga{\gamma}

\def\vare{\varepsilon}

\def\rm{\mathrm}
\def\cal{\mathcal}
\def\scr{\mathscr}

\def\bb{\mathbb}

\def\dexp{\bar{\delta}}

\title{%
Renormalization group and spectra of the generalized Pöschl-Teller potential
}
\date{\today}

\author{
Ulysses Camara da Silva\thanks{ulyssescamara@gmail.com}
\and
Carlos F.S. Pereira\thanks{carlos.f.pereira@edu.ufes.br}
\\
{\small 	Universidade Federal do Esp\'irito Santo} -- 		
{\small		Departamento de Física} \\
{\small	Av. Fernando Ferrari, Goiabeiras, 
		29075-900, 
		Vit\'oria-ES, Brasil}
\and
Andre Alves Lima\thanks{alves.lima@ufpe.br} \\ 
{\small 	Universidade Federal de Pernambuco} --
{\small		Departamento de Física} \\
{\small	Av. Professor Luiz Freire,
		Cidade Universitária, 50670-901,
		Recife-PE, Brasil}
}
\date{\today}

\begin{document}
\pagenumbering{gobble}

\maketitle

\vspace{-5mm}

\begin{abstract}

We study the Pöschl-Teller potential 
${\cal V}(x) = \a^2 g_s \sinh^{-2}(\a x) + \a^2 g_c \cosh^{-2}(\a x)$,
for every value of the dimensionless parameters $g_s$ and $g_c$, including the less usual ranges for which the regular singularity at the origin prevents the Hamiltonian from being self-adjoint. We apply a renormalization procedure to obtain a family of well-defined energy eigenfunctions, and study the associated renormalization group (RG) flow. We find an anomalous length scale that appears by dimensional transmutation, and spontaneously breaks the asymptotic conformal symmetry near the singularity, which is also explicitly broken by the dimensionful parameter $\a$ in the potential. These two competing ways of breaking conformal symmetry give the RG flow a rich structure, with phenomena such as a possible region of walking coupling, massive phases, and non-trivial limits even when the anomalous dimension is absent. We show that supersymmetry of the potential, when present, is also spontaneously broken, along with asymptotic conformal symmetry. We use the family of eigenfunctions to compute the S-matrix in all regions of parameter space, for any value of anomalous scale, and systematically study the poles of the S-matrix to classify all bound, anti-bound and metastable states, including quasi-normal modes. The anomalous scale, as expected, changes the spectra in non-trivial ways.


\end{abstract}

\newpage
\pagenumbering{arabic}

\tableofcontents

\section{Introduction}

The Pöschl-Teller potential is one of the rare exactly solvable potentials in quantum mechanics. 
We can write it in its general form with a convenient parametrization,
\be	\label{V_PT_intro}
V(x) = \frac{2m \alpha^2}{\hbar^2} \left(\frac{g_s}{\sinh^2(\alpha x)}+\frac{g_c}{\cosh^2(\alpha x)}\right)
\ee
where $m$ has units of mass, such that the energy eigenstates of a particle of mass $m$ is determined by the three parameters: $\a$, which sets a  length scale, and $g_s$ and $g_c$, which are dimensionless.
Although the work of Herta Pöschl and Edward Teller \cite{Poschl:1933zz} is now centenary, (\ref{V_PT_intro}) is relevant in many modern applications, most notably in the computation of quasi-normal modes of black holes and other spacetimes
\cite{Ferrari:1984zz,Beyer:1998nu,QNM_dS,Lopez-Ortega:2006aal,cardona_2017_PT,QNM_BH_Julio}.
The potential (\ref{V_PT_intro}) has a regular singular point at $x = 0$, if $g_s \neq 0$.
In applications, when present, this singularity is usually repulsive, and one imposes, accordingly, that the wave function vanishes at the origin. But if one carefully considers, in all generality, \emph{all possible values} of the parameters $g_s$ and $g_c$, the problem of the boundary conditions of (\ref{V_PT_intro}) at the regular singularity becomes subtle: for $g_s < \frac34$, the Hamiltonian is not self-adjoint. 
This is because in the vicinity of $x = 0$ the Pöschl-Teller potential becomes the conformal potential \cite{deAlfaro:1976vlx,Calogero:1970nt}, $V(x) \sim g_s / x^2$, and inherits its well-known properties 
\cite{Renor_Inver_Squa,Camblong:2003mb,Kaplan:2009kr,renor_orig,renor_Group_Limit_Cycle,renor_m_w_c,Gitman}.
The issue with (\ref{V_PT_intro}) is very similar the situation described by Wald and Ishibashi for fluctuations around AdS spacetime \cite{Ishibashi:2004wx}; the equations for fluctuation modes can be reduced to scalar equations with an effective potential like (\ref{V_PT_intro}), but with trigonometric, instead of hyperbolic functions. The term $1/\sin^{2}(\a x)$ then produces an asymtpotic conformal symmetry near $x = 0$, where the boundary of AdS is located. For fluctuations around pure de Sitter space, one finds the hyperbolic potential (\ref{V_PT_intro}), as shown in \cite{QNM_dS}.

In the present paper, we will study the entire parameter space of the generalized Pöschl-Teller potential (\ref{V_PT_intro}), i.e.~all regions of the $g_s$-$g_c$ plane, including the less-known regions where the Hamiltonian is not self-adjoint.
We will show how the energy eigenfunctions can be found from a \emph{renormalization procedure}, analogous to the one developed for the conformal potential \cite{Renor_Inver_Squa,Camblong:2003mb,Kaplan:2009kr,renor_orig,renor_Group_Limit_Cycle,renor_m_w_c}. The result is a one-parameter family of eigenfunctions, with a renormalization group (RG) providing a clear interpretation of the parameter as a length scale $L$ that appears by ``dimensional transmutation'', and breaks the asymptotic conformal symmetry near the singularity. 
The RG flow that we find for the Pöschl-Teller potential has very interesting phenomenology, because \emph{there are two scales competing} for the breaking of asymptotic conformal symmetry. The anomalous scale $L$ induces a \emph{spontaneous breaking}, while the parameter $1/\a$, present in (\ref{V_PT_intro}), induces an \emph{explicit breaking}.
Thus, if the scales are very different, one of them breaks the symmetry ``first'', and the interplay between scales can result in phenomena like a ``walking coupling'' phase, coming from an ``almost fixed point''. 
If, on the other hand, $L \sim 1/\a$, then we have the simultaneous effects of both kinds of breaking, leading to a massive phase to the RG flow.

Spontaneous breaking can be fully prevented if the anomalous scale is set to $L = 0$ or $L = \infty$. The case $L = 0$ corresponds to the solutions usually found in the literature, as it is enforced if the singularity at $x = 0$ is ``sufficiently repulsive''. We will also show that, in some regions of the $g_s$-$g_c$ plane, it is related to the Pöschl-Teller potential being supersymmetric --- thus the anomalous scale also breaks supersymmetry.
We will show that in these two cases where the anomalous scale is absent, one has an interesting feature: there is not one, but a pair of different beta functions, which cannot be perturbatively derived from the beta function of the conformal potential, each describing a different type of massive RG flow with a UV fixed point and a mass scale set by $\a$.

Once we have the energy eigenfunctions parameterized by the anomalous dimension, we proceed to investigate the spectra of the Pöschl-Teller Hamiltonian, again in every region of its coupling space. We compute the S-matrix over the entire coupling plane and then, from a systematic analysis of its poles, determine the bound, anti-bound and metastable states. We thus construct three ``phase diagrams'', Figs.\ref{Moduli_L_theta}, \ref{Moduli_IR} and \ref{Moduli_L_theta_UV},
giving the spectra in the $g_c$-$g_s$ plane for $L$ finite, $L =0$ and $L = \infty$.
As expected, there are qualitatively different spectra depending on whether the anomalous scale is finite, or zero, or infinite, and we find some novel phenomena. For instance, if the potential is strongly attractive near the origin, and the asymptotic conformal symmetry is spontaneously broken, the spectrum of bound states displays something similar to the Efimov effect \cite{Efimov}, which is a property of the conformal potential, but with a perceptible correction due to the \emph{explicit} breaking of conformal symmetry by the scale $1/\a$.
On the other hand, if we set the anomalous scale $L = 0$, forcing the wave function to vanish at $x = 0$, the metastable states satisfy the boundary conditions of quasinormal modes (QNM) for the singular Pöschl-Teller potential, see \cite{QNM_dS,Lopez-Ortega:2006aal}. Here we classify these quasinormal modes, along with the bound states, for \emph{all} values of the parameters $g_c$-$g_s$.

\bigskip

The structure of this paper is as follows. 
In Sect.\ref{SectSolofSch} we describe the coupling plane $g_c$-$g_s$ with its qualitatively different regions, and the problem of boundary conditions at the singularity. We examine in which regions the Pöschl-Teller potential is supersymmetric, and how SUSY and shape-invariance can be used to fix the ill-defined boundary conditions for certain ranges of the couplings.
In Sect.\ref{sec:renor}, we first apply the renormalization procedure to obtain the general family of energy eigenfunctions, parameterized by an anomalous scale.
Then, we define the renormalization group with a running coupling and the associated beta function. In this framework, we study how the asymptotic conformal symmetry of the Pöschl-Teller potential is broken, either spontaneously, by the anomalous dimension, or explicitly, by the length scale $1/\a$.
In Sect.\ref{SectSpectra}, we compute the S-matrix for the potential in all generality, and systematically search for bound, anti-bound and metastable states, and instabilities in all regions of parameter space, including new types of bound states with a kind of ``Efimov effect'' in \S\ref{SectSpectLthe}, and quasinormal modes in \S\ref{SectSpectL0}.
In Sect.\ref{SectConclusion} we make our concluding remarks.
Details of some computations are given in Appendices \ref{ApednSwhofom} and \ref{appen}.

\section{Boundary conditions and parameter space}\label{SectSolofSch} 

\begin{figure}[t] 
\centering
\subfigure[]{
\includegraphics[scale=0.35]{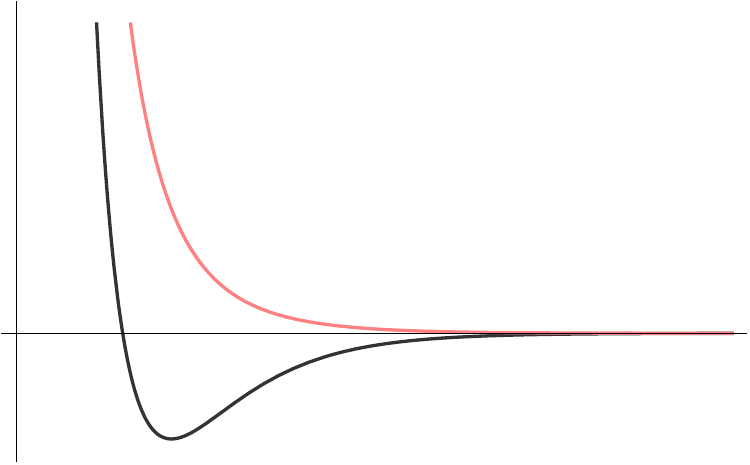}
\label{fig:V_1}
}
\subfigure[]{
\includegraphics[scale=0.35]{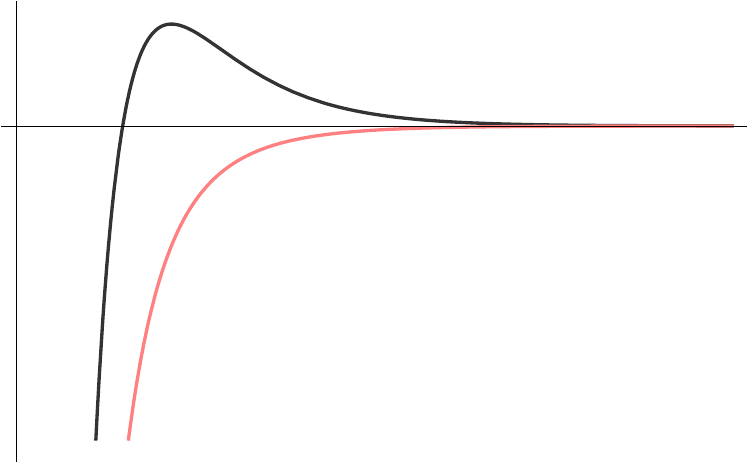} \label{gif:V_2}
}
\subfigure[]{
\includegraphics[scale=0.35]{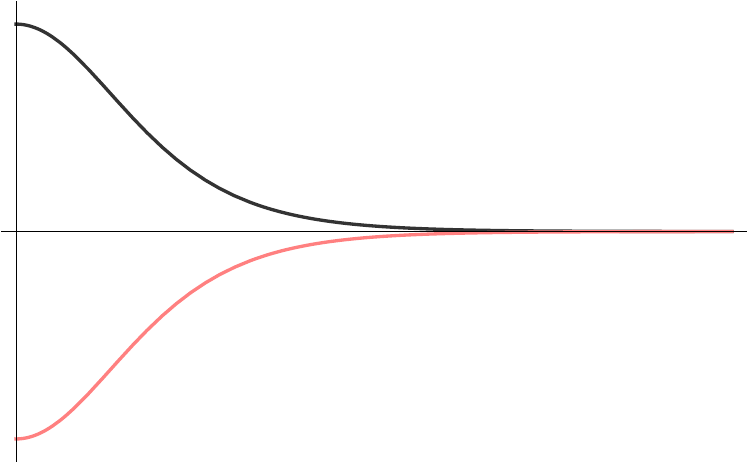}
\label{fig:V_3}
}
\caption{Shapes of ${\cal V}(x)$. \\
(a) $g_s>0$. Black: $g_c<0$ and $g_s/|g_c|<1$. Red: $g_s/|g_c|>1$. \\
(b) $g_s<0$. Black: $g_c>0$ and $|g_s|/g_c<1$. Red: $|g_s|/g_c>1$. \\
(c) $g_s=0$. Black: $g_c>0$. Red line: $g_c<0$.
}
\label{fig:V_PT}
\end{figure}

Let us define the rescaled Pöschl-Teller potential
\be	\label{V_PT}
\mathcal{V}(x)=\alpha^2\left(\frac{g_s}{\sinh^2(\alpha x)}+\frac{g_c}{\cosh^2(\alpha x)}\right), 
\qquad
 0<x<\infty,
\ee
related to (\ref{V_PT_intro}) by ${\cal V}(x) = 2m \hbar^{-2}V(x)$.
Here $\a > 0$ has dimension of inverse length and does not affect the shape of ${\cal V}$. Our concern will be the two dimensionless couplings $g_s$ and $g_c$, which define a two-dimensional `coupling space'. As we will show, in different regions of the coupling plane, the properties of the potential and its spectrum can be very distinct.
First of all, ${\cal V}(x)$ has qualitatively distinct shapes, depending on how the couplings $g_c$ and $g_s$ are chosen, as shown in Fig.\ref{fig:V_PT},
from which it is clear that different shapes of ${\cal V}(x)$ may support or not bound states.

We have chosen the normalization of (\ref{V_PT}) such that the stationary Schrödinger equation reads
\be	\label{Sch_x}
\frac{d^2\psi(x)}{dx^2} + \big[ k^2- {\cal V}(x) \big] \psi(x)=0, 
\qquad 
k = \sqrt{2mE} / \hbar .
\ee
For now, we will take $k\in {\bb C}$ to be arbitrary, and consider specific values of interest later. 
The goal is to first find the general solution of Eq.(\ref{Sch_x}). 
Defining $u=\tanh^2(\alpha x)$, $0<u<1$, the Schrödinger equation becomes
\be	\label{EDO_u}
u (1-u) \frac{d^2\psi(u)}{du^2} + \left( \frac{1}{2}-\frac{3}{2}u \right) \frac{d\psi(u)}{du} + \left(\frac{1}{(1-u)}\frac{k^2}{4\alpha^2}-\frac{1}{u}\frac{g_s}{4}-\frac{g_c}{4}\right) \psi(u) = 0 .
\ee
The general solution can be written as
\bsub	\label{sol_u}
\be
\psi_k(x) 
	= \big[ \tanh(\alpha x) \big]^{\frac{1}{2}+\nu}
	\big[ \cosh(\alpha x) \big]^{- i \frac{k}{\alpha}}
	F(x),
\ee
where $F(x)$ is a combination of hypergeometric functions,
\begin{align}
\begin{aligned}
&F(x) 
	= \frac{A_k}{\a^{\frac{1}{2}+\nu}} \,
	{}_2F_1\left[
		\tfrac12 (1 + \n - \m + ik/\a) ,
		\tfrac12 (1 + \n + \m + ik/\a);
		1+\nu;
		\tanh^2(\a x)
		\right]
	\\
	&
	+ 
	\frac{B_k}{\a^{\frac{1}{2}-\nu}}
	\frac1{\tanh^{2\n} (\a x)} 
	\,{}_2F_1 \left[
		\tfrac12 (1 - \n - \m + ik/\a) ,
		\tfrac12 (1 - \n + \m + ik/\a);
		1 - \n;
		\tanh^2(\a x)
		\right]
\end{aligned}
\end{align}\esub
and we have defined  
\be	\label{munudefA}
\nu = \sqrt{\frac{1}{4}+g_s}, \qquad \mu = \sqrt{\frac{1}{4}-g_c}.
\ee
The solution (\ref{sol_u}) is not valid if $\nu\in\mathbb{Z}_{\ge0}$, but after analyzing boundary conditions at $x = 0$ we find that only non-trivial case is when $\nu = 0$, i.e.~$g_s=-1/4$. In this case there is a logarithmic branch cut, as discussed in Appendix \ref{appen}. Unless specified, we will assume that $\nu \neq 0$.

\subsection{The regions of coupling space}	\label{SectRegionCoup}

The parameters $\m$, $\n$ defined in (\ref{munudefA}) can be used instead of $g_s$ and $g_c$ to describe the coupling space,
\be	\label{munudef}
\begin{aligned}
\nu &= \sqrt{\tfrac{1}{4}+g_s} , \\
\mu &= \sqrt{\tfrac{1}{4}-g_c} ,
\end{aligned}
\qquad\qquad
\begin{aligned}
g_s &= - \tfrac14 + \n^2 ,\\
g_c &= + \tfrac14 - \m^2 .
\end{aligned}
\ee
In fact, many of the features that we will explore are more naturally described in terms of $\m$ and $\n$, instead of $g_c$ and $g_s$.

\begin{figure}[t] 
\centering
\includegraphics[scale=0.32]{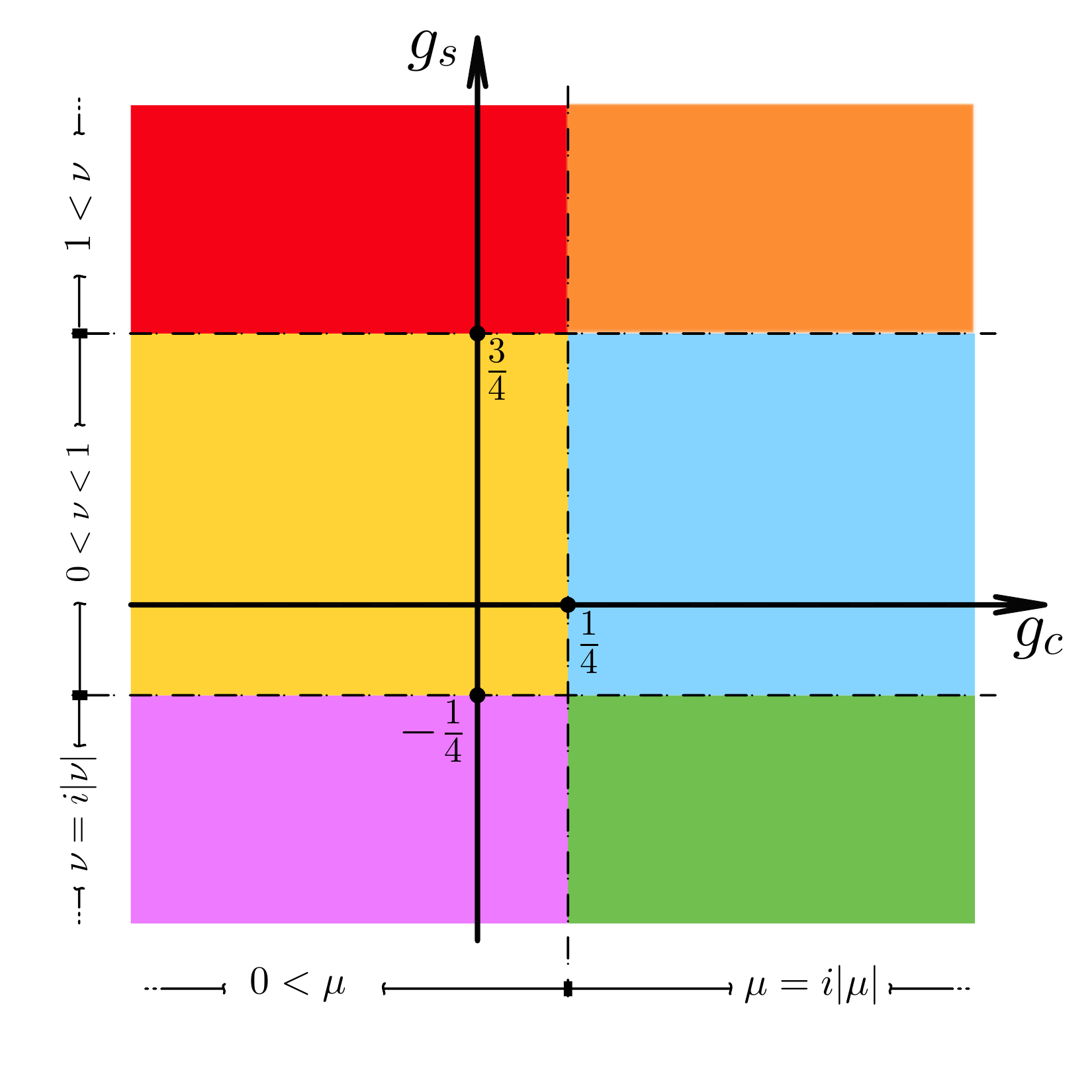}
\caption{%
Coupling space, and its regions.
Red and orange patches: strongly repulsive region.
Yellow and blue: weak medium.
Purple and green: strongly attractive.
The vertical line $g_c = \frac14$ marks the division between real and imaginary $\m$.
}
\label{Moduli_plane}
\end{figure}

The parameters $\m$ and $\n$ relate directly to properties of the eigenfunctions. Their behavior near the singularity is controlled by the value of $\n$, and behavior at infinity by the value of $\m$. 
Either one of $\nu$ and $\mu$ may be real or purely imaginary, and the lines $g_s = - \frac14$ and $g_c = \frac14$, where they change from real to imaginary define the boundaries of regions in the coupling plane where the Pöschl-Teller system changes qualitatively. In the end, the coupling plane can be divided into six semi-infinite regions, represented in Fig.\ref{Moduli_plane}.

The tree different regions across the horizontal $g_s$-axis in Fig.\ref{Moduli_plane} correspond to the three possible behaviors of the solutions near the singularity.
The constants $A$ and $B$ in (\ref{sol_u}) where chosen such that the asymptotic solution near the origin looks simple. 
Expanding the hypergeometrics for $\alpha x\ll 1$, we find that,
if $g_s > -\frac{1}{4}$, hence $\n$ is real, then
\bsub	\label{x_0_both}
\be	\label{x_0}
\psi_k (x) \approx x^{1/2}\left[A_k \, x^{\nu}+B_k \, x^{-\nu}\right] .
\ee
This is the behavior in the red, orange, yellow and blue patches of Fig.\ref{Moduli_plane}.
On the remaining purple and green patches, $g_s < - \frac14$, hence $\nu = i |\nu|$ is imaginary, and
\be	\label{x_<}
\psi_k (x) \approx x^{1/2} \left[ A_k \, x^{i|\nu|} + B_k \, x^{-i|\nu|} \right] ,
\qquad
A_k =B_k^*
\ee
\esub
The constants $A$ and $B$ must be complex conjugates, so that $\psi_k(x)$ is real, since this is the solution of a real ODE. 

The solution must be square-integrable near the singular point,
\be	\label{SquaInteg}
\lim_{x \to 0} \int^x |\psi_k(x')|^2 dx' < \infty.
\ee
From the asymptotic forms (\ref{x_0_both}), we find three distinct possibilities. 
\begin{enumerate}[\itshape i)]
\item	\label{RenItem1}
If $g_s \geq \frac34$, hence $\nu \geq 1$, condition (\ref{SquaInteg}) fixes $B = 0$ in Eq.(\ref{x_0}). In parameter space, this region is made by the (infinite) red and orange patches shown in Fig.\ref{Moduli_plane}.
Following nomenclature from the conformal potential literature, we will say that these patches form the ``strongly repulsive region''.

\item \label{RenItem2}
If $- \frac14 \leq g_s <\frac{3}{4}$, hence $0 \leq \nu < 1$, both $x^\nu$ and $x^{-\nu}$ are square-integrable. Then condition (\ref{SquaInteg}) does \emph{not} fix any of the constants $A$ or $B$ in Eq.(\ref{x_0}). 
These conditions are met at the yellow and blue patches of Fig.\ref{Moduli_plane}, which we will call the ``weak medium'' region in coupling space.

\item \label{RenItem3}
If $g_s < - \frac14$, hence $\nu = i |\nu|$, the solution is (\ref{x_<}), which is strongly oscillating as $x \to 0$.
This holds in the purple and green patches of Fig.\ref{Moduli_plane}, which form the ``strongly attractive region'' of coupling space.

\end{enumerate}
The ambiguity in boundary conditions in cases \ref{RenItem2} and \ref{RenItem3} mean that the conformal Hamiltonian is not self-adjoint for $g_s < \frac34$ \cite{Gitman}.

\subsection{Supersymmetry and boundary conditions in the weak-medium region}	\label{bound_c}

Two Hamiltonians $H_\pm$ are a `supersymmetric' pair if they can be factorized as a products of the conjugate operators%
	\footnote{We use units were $\hbar=\sqrt{2m}=1$}
\be	\label{QQbar}
Q = \frac{d}{dx} + W(x), 
\qquad
Q^{\dagger}=-\frac{d}{dx} + W(x) ,
\ee
defined by a (real) `superpotential' function $W(x)$, such that  \cite{Infeld:1951mw,Cooper:1994eh}
\be	\label{SUSYVs}
\begin{aligned}
H_+ &= Q^\dagger Q
	= - d^2 / dx^2 + U_+(x)
\\
H_- &= Q Q^\dagger
	= - d^2 / dx^2 + U_-(x)
\end{aligned}
\qquad
\left[
\begin{aligned}
U_+(x) &= W^2(x) - W'(x)  \\
U_-(x) &= W^2(x) + W'(x) 
\end{aligned}
\right.
\ee
The eigenstates $\psi_k^\pm$ of $H_+$ and $H_-$ are said to be ``bosonic'' and ``fermionic'', respectively. In our units, we can write $H_\pm \psi^\pm_k = k_\pm^2 \psi_k^\pm$. 
Supersymmetry (SUSY) is said to be ``spontaneously broken'' if there is no bosonic zero-mode with $E^+_0 = 0$, i.e.~if $\psi^+_0(x)$ is \emph{not} normalizable. In this case, the bosonic and fermionic spectra coincide exactly,%
	\footnote{
	If the bosonic zero-mode $\psi^+_0$ \emph{is} normalizable, i.e.~if there is a state with $E^+_0 = 0$, then SUSY relates the bosonic and fermionic spectra $\{ E^\pm_n\}_{n=0}^\infty$ as $E^-_n = E^+_{n+1}$.
	}
$k_+ = k_- = k$, and the eigenstates are related by
\be	\label{SUSYpartpsi}
\psi^-_k (x) = k^{-1} Q \psi^+_k(x) , 
\qquad
\psi^+_k(x) = k^{-1} Q^\dagger \psi^-_k(x) .
\ee

If the two couplings, $g_s$ and $g_c$, of the Pöschl-Teller potential lie in the range
\be
g_s \geq - \tfrac14 , \qquad g_c \leq \tfrac14 ,
\ee
then the two parameters defined in (\ref{munudef}) are \emph{real} and positive,
$\mu \geq 0$, $\nu \geq 0$.
In this case, the potential (\ref{V_PT}) is supersymmetric, because we can define the superpotential \cite{dutt1988}
\be	\label{W}
W(x) = -\alpha \left( \nu + \tfrac{1}{2} \right) \coth(\alpha x)
	+ \alpha \left( \mu - \tfrac{1}{2}\right) \tanh(\alpha x) ,
\ee
in terms of which the Pöschl-Teller potential ${\cal V}(x)$ has the form (\ref{SUSYVs}), apart from a shift in the energy:
$U_+(x) = {\cal V}(x) - \a^2 (\m - \n -1 )^2 .$
So the Pöschl-Teller Hamiltonian can be written as the ``bosonic'' partner of a SUSY pair, 
\bsub
\begin{align}
H^\rm{PT}_+ - \alpha^2\left(\mu-\nu-1\right)^2 &= Q^{\dagger} Q , \\
H^\rm{PT}_- - \alpha^2\left(\mu-\nu-1\right)^2 &= Q Q^{\dagger} ,
\end{align}
\esub
with the operators $Q$ and $Q^\dagger$ given in (\ref{QQbar}) by the superpotential (\ref{W}).
In fact, the ``fermionic'' partner $H^\rm{PT}_-$ is, \emph{again}, a Pöschl-Teller Hamiltonian. That is, the Pöschl-Teller potential is \emph{`shape-invariant'} \cite{Cooper:1994eh}; the pair of partner potentials are
\be	\label{WminusW}
\begin{aligned}
U_+(x) 
	&\equiv W^2(x) - W'(x) 
\\
	&= 
	\alpha^2 \left[
	\frac{\nu^2 - \frac{1}{4}}{\sinh^2(\alpha x)}
	+ \frac{ \frac{1}{4} - \m^2 }{\cosh^2(\alpha x)}
	+\left(\mu-\nu-1\right)^2
	\right]
\\
U_-(x) 
	&\equiv W^2(x) + W'(x)
\\
	&=
	\alpha^2 \left[
	\frac{(\nu+1)^2-\frac{1}{4}}{\sinh^2(\alpha x)}
	+ \frac{\frac14 - (\mu-1)^2 }{\cosh^2(\alpha x)}
	+ (\mu-\nu-1)^2
	\right]
\end{aligned}	
\ee
and one can see that $U_+(x)$ and $U_-(x)$ ``have the same shape'', i.e.~they are the same function of $x$, only with shifted parameters.
More precisely, discounting the energy shift, we have the pair of Pöschl-Teller potentials 
\be
\begin{aligned}
{\cal V}_+(x; \m,\n) 
	&\equiv U_+(x) - \alpha^2 (\mu-\nu-1)^2 
	= 	
	\alpha^2 \left[
	\frac{\nu^2 - \frac{1}{4}}{\sinh^2(\alpha x)}
	+ \frac{ \frac14 - \mu^2 }{\cosh^2(\alpha x)}
	\right]
\\
{\cal V}_-(x; \tilde\m , \tilde\n) 
	&\equiv U_-(x) - \alpha^2 (\mu-\nu-1)^2 
=	
	\alpha^2 \left[
	\frac{\tilde\nu^2 - \frac{1}{4}}{\sinh^2(\alpha x)}
	+ \frac{\frac14 - \tilde\mu^2 }{\cosh^2(\alpha x)}
	\right]
\end{aligned}	
\ee
which only differ by the shifted parameters
\be	\label{partnarem}
\nu \mapsto \tilde \nu = \nu +1  ,
\qquad
\mu \mapsto \tilde\mu = \mu - 1 .
\ee

SUSY and shape-invariance give us a criterion to fix the boundary condition of the eigenfunctions at the singularity in the part of the weak-medium region of coupling space where $0 < \nu < 1$ is real (yellow patch of Fig.\ref{Moduli_plane}).
If we assume that the bosonic potential ${\cal V}_+(x)$ is in this region, then (\ref{partnarem}) shows that its fermionic partner ${\cal V}_-(x)$ will be in the strongly repulsive region, where $\tilde g_s$ has $\tilde\nu > 1$ (red patch of Fig.\ref{Moduli_plane}).
In the strongly repulsive region, the zero-energy wave function is not normalizable, so SUSY is spontaneously broken, and we can relate every strongly-repulsive eigenstate $\psi_k^-(x)$, which are well-defined, to a weak-medium eigenstate $\psi^+_k(x)$, via Eqs.(\ref{SUSYpartpsi}). Hence the well-defined functions in the red patch fix the boundary conditions of the otherwise ill-defined functions in the yellow patch.

In order to make it very clear how imposing SUSY fixes the boundary conditions in the weak-medium region, recall that, in this region, the most general solution behaves for $x \to 0$ as
\be	\label{x_0bis}
\psi_k^+ (x) \approx A_k\, x^{\n + \frac12 } + B_k\,  x^{-\nu + \frac12} ,
\qquad
0 < \n < 1 .
\ee
This function is square-integrable at $x = 0$ for any choice of $A_k$ or $B_k$.
From Eq.(\ref{SUSYpartpsi}), the partner state is
\be
\begin{aligned}
\psi^-_k(x) &= k^{-1} Q \psi^+_k(x) 
	\approx c_1 A_k \, x^{\n - 1 + \frac12 } + c_2 B_k \, x^{-\nu - 1 + \frac12} ,
\qquad
0 < \n < 1 .
\end{aligned}
\ee
with $k$-dependent constants $c_1$, $c_2$. 
\emph{This} latter function is only square-integrable if 
$B_k = 0 .$
So, in short, although generic functions (\ref{x_0bis}) are always square-integrable in the weak-medium range, only the functions
\be	\label{x_0bisFix}
\psi_k^+ (x) \approx A_k\, x^{\n + \frac12 } ,
\qquad
0 < \n < 1 ,
\ee
match the supersymmetric relation between the strongly-repulsive and the medium-weak regions of coupling space.
Hence, to enforce the naturally existing supersymmetry of the Hamiltonians of these two regions, we must fix the (otherwise arbitrary) boundary conditions in the medium-weak region as in (\ref{x_0bisFix}).

\section{The renormalization group of the Pöschl-Teller potential}\label{sec:renor}

One way of making the boundary conditions well-defined is to perform a \emph{renormalization} of the potential near the singularity. Renormalization of the conformal potential is well-known \cite{Renor_Inver_Squa,Camblong:2003mb,Kaplan:2009kr,renor_orig,renor_Group_Limit_Cycle,renor_m_w_c}; the result is technically equivalent to the construction of a self-adjoint extension of the Hamiltonian \cite{Gitman}, but with the advantage of providing a clear physical interpretation to the parameters involved.

\subsection{Renormalization}	\label{SectrenorSub}

To perform the renormalization of the Pöschl-Teller potential, we adopt a general approach that does not require, a priori, imposing boundary conditions at the origin, nor defining an explicit regulating potential, and, most importantly, that can be used for any potential with an inverse-square singularity. 
The first step is to introduce an arbitrary cutoff $R$, below which we replace the singular part of ${\cal V}(x)$ by an arbitrary function $f(x/R)$,
\be
{\cal V}_R (x) =
	 \left\{ 
	 \begin{aligned}
	&{\cal V}(x) && x>R \\
	&\frac{\lambda(R)}{R^2}f(x/R) \qquad && 0<x<R
	\end{aligned}
	\right.
\ee
We only require that $f(x)$ is not singular at $x = 0$. The parameter $\la(R)$ (which does not depend on $x$) is introduced  so that $f(1) = 1$.
Usually, $f$ is chosen to be a step function \cite{renor_orig}, but it is not necessary to specify it at all.
The regularization procedure can be thought of in terms of an `effective' description of the singularity, see e.g.~\cite{Camblong:2003mb}.
We look for solutions outside the ``core region'' $x < R$, and assume that, at energies $k$ that are small compared with the energy scale $1/R$, i.e.~for 
\be	\label{kR11}
k R \ll 1,
\ee
physics must be insensitive to what happens inside in the regularized core.

Let us denote the ``unphysical'' wave function inside the regularized region $x < R$ by $\psi_{k}^{<}(x)$. Eq.(\ref{Sch_x}) reads
\be	\label{SchRens}
\frac{d^2\psi_{k}^{<}(x)}{dx^2}+\left(\frac{(kR)^2-\lambda(R)f(x/R)}{R^2}\right)\psi_{k}^{<}(x)=0 .
\ee
At energies for which the effective description is valid, $kR \ll 1$, we can neglect the term $(kR)^2$, and Eq.(\ref{SchRens}) becomes independent of $k$. 
The solution $\psi_{k}^{<}$ for $k \neq 0$ is then the same as the solution $\psi_{0}^{<}$ for $k = 0$. By continuity, so are their logarithmic derivatives,
\be	\label{CalFdefPrev}
\psi'^<_{k} (R) / \psi_{k}^<(R)
	=  \psi'^<_0(R) / \psi_{0}^{<}(R) 
	\equiv {\cal F}(R),
\ee
where ${\cal F}(R)$, by definition, does not depend on $k$. 
The wave function and its first derivative must be continuous for all $x$, so we conclude that
\be	\label{CalFdef}
{\cal F}(R) = \frac{\psi'_0(R)}{\psi_{0}(R)} ,
\ee
where $\psi_0(x)$ is the ``physical'' solution \emph{outside} the regularized region, and, more generally, 
\be	\label{Robin}
\frac{\psi'_{k}(R)}{\psi_{k}(R)} = \mathcal{F}(R) 
\qquad \text{for} \qquad kR \ll 1 .
\ee
So \emph{the renormalization procedure amounts to imposing Robin boundary conditions at the regulating point} $x = R$, without specifying any conditions at the singular point $x = 0$. 

It is important to emphasize that the function ${\cal F}(R)$ is \emph{defined} by Eq.(\ref{CalFdefPrev}), 
hence it is completely (although indirectly) fixed by the ``strength'' of the potential inside the regularizing core,%
	\footnote{%
	This point of view, of making the amplitude of the regularizing potential change with $R$,  is analogous to the ``core renormarlization framework'' of \cite{Camblong:2003mb}.}
 i.e.~by $\la(R) f(x/R)$, through the logarithmic derivative of the function $\psi^<_0(x)$. 
That is to say, ${\cal F}(R)$ does not depend in any way on what happens outside the regularizing core. In particular, it does not depend on the form of the pair of independent solutions of the Pöschl-Teller Schrödinger equation (\ref{Sch_x}).
The meaning of Eq.(\ref{Robin}) is that, for a given value of $R$, the number ${\cal F}(R)$ --- which is independent of the Schrödinger equation --- \emph{fixes the boundary conditions} of the otherwise general solution (\ref{sol_u}) of Eq.(\ref{Sch_x}). In other words, ${\cal F}(R)$ defines a relation between the integration constants $A_k$ and $B_k$.

We can find this relation as follows. 
First, it is convenient to define the dimensionless quantity
\be	\label{gamma}
\gamma(R) \equiv - \frac{1}{2} +  R {\cal F}(R) ,
\ee
and work with $\ga(R)$ instead of ${\cal F}(R)$. Now, note that we can always compute ${\cal F}(R)$, and hence $\ga(R)$, from the zero-energy solution of the Pöschl-Teller potential, since when $k = 0$ condition (\ref{kR11}) is identically satisfied. 
The result for $\ga (R)$ can be organized as a ratio of two series in powers of $(\a R)$, see Appendix \ref{ApednSwhofom}.
Since $1/\a$ is the characteristic length scale of the Pöschl-Teller potential, we must take
\be	\label{alphaR1}
\a R \ll 1.
\ee
(Otherwise the potential would become too disfigured by a large regularized region.) Thus we find
\be	\label{gaRa2}
\ga(R) 
	=
	\n
	\frac{1 - \vare (L/R)^{2\n}}{1 + \vare (L/R)^{2\n}}
\ee
ignoring terms of order $\a R$.
Here 
\be	\label{varedef}
\vare \equiv \rm{Sign} [B_0 / A_0] = \pm 1,
\ee
and we have defined
\be	\label{ren_g_int0}
L^{2\nu} \equiv | B_0 / A_0 | ,
\ee
which is an intrinsic and \emph{arbitrary} length scale contained in the solution $\psi_0(x)$ given by Eq.(\ref{sol_u}).
We are assuming for now that $\n > 0$ is real. 
If we compute the equivalent of the r.h.s.~of Eq.(\ref{gamma}), in the effective regime (\ref{kR11}), and taking (\ref{alphaR1}) and (\ref{Robin}) into account, the result must coincide with the r.h.s.~of Eq.(\ref{gaRa2}). 
(See Eq.(\ref{almsGrakMai}) in Appendix \ref{ApednSwhofom}.)
As a consequence,
\be	\label{ren_g_int}
B_k / A_k  = \vare L^{2\nu} ,
\ee
which holds for all $k$ when we take $R \to 0$.
In other words, the ratio of the integration constants $B_k / A_k$ does not depend on $k$.

\subsection{Breaking of asymptotic conformal symmetry}	\label{SectRenorStrontehta}

We can associate a renormalization group (RG) flow to the renormalization above.
We interpret $\ga(R)$ as a `running coupling'.%
	\footnote{NB: When we refer to `coupling space', we always mean only the couplings $g_s$, $g_c$ of the Pöschl-Teller potential.}
(In fact, the specific form of Eq.(\ref{gamma}) is in fact inspired by the RG of the conformal potential, see e.g.~\cite{renor_m_w_c,U_AB,Lima:2019xzg}.)
The flow of $\ga(R)$ is characterized by its beta function
\be	\label{betaDef}
\beta(\ga) \equiv \frac{d \ga}{d \log R} .
\ee
Now, note that, if we take the derivative of 
$\ga(x) = - \frac12 + x \psi_0'(x) / \psi_0(x)$, 
use the Schrödinger equation, then evaluate the result at $x = R$, we find 
\be	\label{betaiwV}
\beta(\ga) = - \ga^2 + \tfrac14 + R^2 {\cal V}(R).
\ee
Expanding ${\cal V}(x)$ up to first order in $x$, Eq.(\ref{betaiwV}) gives
\be	\label{betaiwVEx}
\beta(\ga) =  -(\ga + \n)(\ga - \n) - \tfrac13  ( \n^2 + 3\m^2 -1) (\a R)^2  + \rm{O}(\a R)^4
\ee
Here it is understood that we must find $R = R(\ga)$, to eliminate $R$ from the r.h.s.~in favor of $\ga$, by inverting Eq.(\ref{gaRa3}). 
As shown in Appendix \ref{ApednSwhofom}, this is difficult to do analytically, even at lowest order, because of the terms $(L / R)^{2\n}$ in the asymptotic expansion of $\ga(R)$.

\subsubsection{Spontaneous and explicit symmetry breaking}	\label{SextSpontBreakExpl}

\emph{Near the singularity}, if we make $\a R \to 0$ in Eq.(\ref{betaiwVEx}), we find
\be	\label{beta1}
\beta (\ga) = - (\ga + \n)(\ga - \n) + \cdots
\ee
which is consistent with $\ga(R)$ given by Eq.(\ref{gaRa2}).
The dots indicate terms that we have neglected when taking $\a R = 0$.
Eq.(\ref{beta1}) is the beta-function of the conformal potential \cite{Kaplan:2009kr}, as expected, since we are in the asymptotic region where the Pöschl-Teller becomes conformal.
The RG flow driven by (\ref{beta1}) has two fixed points, where $\beta (\ga) = 0$,
\be	\label{Fixptsga}
\ga_\rm{IR} = + \nu ,\qquad \ga_\rm{UV} = - \nu .
\ee
Although we will call them ``fixed points'', it should be understood that (\ref{beta1}) is \emph{not}, of course, the exact beta function for the Pöschl-Teller potential; one can think of (\ref{beta1}) as ``the beta function of the asymptotic conformal symmetry'', and $\ga = \pm \n$ as ``the fixed points of asymptotic conformal symmetry''.
 $\ga = + \n$ is an IR fixed point, in the sense that it is reached at ``long distances'', 
\be	\label{IRlimit}
R / L \gg 1;
\ee
while $\ga = - \n$ is a UV fixed point, reached at ``small distances'', 
\be	\label{UVlimit}
R/L \ll 1.
\ee
Thus $L$ causes a \emph{spontaneous breaking} of the 
``classical'' asymptotic conformal symmetry of the Pöschl-Teller potential.
The same phenomenon is found in the exact conformal potential \cite{Kaplan:2009kr,renor_m_w_c,U_AB,Lima:2019xzg}.
But, while the inverse-square potential has no intrinsic length scale, the Pöschl-Teller potential has $1/\a$, which causes an \emph{explicit breaking} of the asymptotic conformal symmetry.

\begin{figure}[t] 
\centering
\includegraphics[scale=0.3]{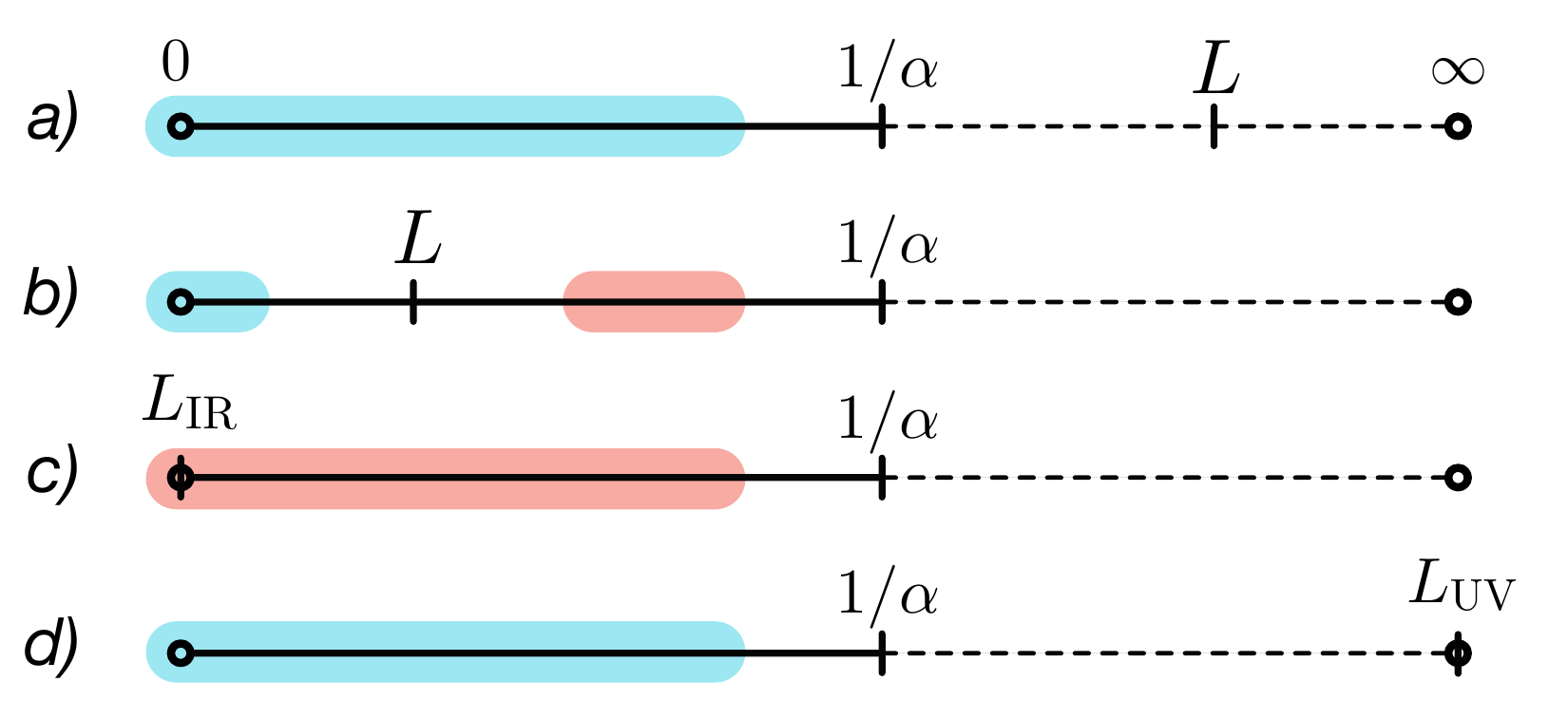}
\caption{
Ranges of the cutoff $R$, depending on the position of the anomalous scale $L$. Red indicates a IR regime (\ref{IRlimit}), and blue a UV regime (\ref{UVlimit}).
}
\label{R_ranges}
\end{figure}

We have a hierarchy of scales: depending on whether $\a L\lessgtr 1$, conformal symmetry may be broken either spontaneously or explicitly first.
We can look at the limits (\ref{IRlimit}) and (\ref{UVlimit}) in units of $1/\a$, as 
$
\a R / \a L .
$
If $\a L \geq 1$, then we are of course forced to stay inside the UV region, as in Fig.\ref{R_ranges}a, and the symmetry is explicitly broken before $L$ plays its part.
But if $L$ is a ``quantum'' scale, in the sense that $\a L \ll 1$, we can slide the cutoff from the UV deep into the IR region, while keeping $\a R \ll 1$, as in Fig.\ref{R_ranges}b, so $L$ spontaneously breaks the asymptotic symmetry before $1/\a$ does.
More precisely, in this latter case, the nature of the asymptotic RG flow depends on the sign $\vare = \pm1$.
Looking at $\ga(R)$ in Eq.(\ref{gaRa2}), we see a ``massless RG flow'' for  $\vare = +1$, with the scale $R/L$ free to run across the UV and IR fixed regions. 
If $\vare = - 1$, on the other hand, $L$ behaves like a ``QCD scale'', where the coupling $\ga(R)$ diverges, and the flow ends in a massive phase with mass scale $1/L$.

\bigskip

In any case, we emphasize that \emph{if $L$ is finite, it is always impossible to actually reach the IR fixed point,} since this requires that $R \to \infty$ in (\ref{gaRa2}). Thus the RG flow always has a massive phase, with mass scale given by $\a$.
What may happen is that, if $\a L$ is very small, but not zero, the asymptotic beta function (\ref{beta1}) will hold for a long while as $\ga(R)$ leaves the UV fixed point $\ga_\rm{UV} = -\n$; then, as $\ga(R) \to + \nu$, the running coupling may start to ``walk''. But eventually the $\a R$ corrections become relevant, and the flow becomes massive.
Finding the effect of $\a$ and its explicit symmetry breaking in the beta function is complicated because it requires inverting a series for $\ga(R)$ and, if $L$ is finite, this series has a complicated structure, as discussed in Appendix \ref{ApednSwhofom}, see Eq.(\ref{gaRa3}).

\subsubsection{RG flows without spontaneous symmetry breaking}

There are two possible ways of getting rid of the anomalous dimension: by taking
\be	\label{UVIRL}
L_\rm{IR} = 0 
\qquad \text{or} \qquad
L_\rm{UV} = \infty.
\ee
If we fix $L = L_\rm{IR}$, the IR regime (\ref{UVlimit}) extends to all values of $R$, as in Fig\ref{R_ranges}c.
Fixing $L = L_\rm{UV}$ instead, it is the UV regime (\ref{IRlimit}) that extends to all values of $R$, as in Fig.\ref{R_ranges}d.
In both cases, the conformal coupling (\ref{gaRa2}) stops running; for $L = \infty$, it stays fixed at the UV fixed point $\ga_\rm{UV} = - \n$, and, for $L = 0$, it stays at the IR fixed point $\ga_\rm{IR} = \nu$. Note that this latter case is the only way of actually reaching the IR fixed point of the conformal beta function (\ref{beta1}). This is an important fact: the IR fixed point $\ga_\rm{IR} = \nu$ only becomes accessible if $L = 0$ \emph{exactly}; otherwise, it is always hidden above the scale $1/\a$.

This ``discontinuity'' --- the fact that $\ga_\rm{IR} = \nu$ is only accessible ``non-perturbatively'' (in the value of $L$) --- is associated with a drastic change in the RG flow when $L$ is set to one of the points (\ref{UVIRL}).
Conformal symmetry is now necessarily broken explicitly, by the scale $1/\a$, since there is no anomalous scale to break it spontaneously. To find the beta function, we must take $\a R$ terms into account in computing the running coupling $\ga(R)$, hence we cannot use neither (\ref{gaRa2}) nor (\ref{beta1}), which discarded $\a R$ terms completely.
We have to compute both $\ga(R)$ and $\beta(\ga)$ from the start, in each case (\ref{UVIRL}).
Fortunately, here it \emph{is} possible to compute perturbatively both functions, as the structure of the $\a R$ series completely changes when we eliminate $L$, as discussed in Appendix \ref{ApednSwhofom}.

In fact, we find two \emph{distinct} RG flows, one going out of each point $\ga = \pm \n$. 
Note that, the conformal beta function (\ref{beta1}) is not a function anymore, but instead a ``beta constant'', $\b(\pm\nu) = 0$. It ceases to be constant only when we find the $\a$ corrections.
Computing $\ga(R)$ from $\psi_0(x)$, using (\ref{ren_g_int0}) to fix the constants $A_k$, $B_k$, we find, at leading order, 
\begin{flalign}
&\text{for $L = L_\rm{IR}$:}	\label{gaRaIR}
&&
\ga(R) = + \n  - \frac{\n^2 + 3\m^2 -1}{6 (1+\n)}  (\a R)^2 + \cdots
&&
\\
&\text{for $L = L_\rm{UV}$:}	\label{gaRaUV}
&&
\ga(R) = - \n - \frac{\n^2 + 3\m^2 -1}{6 (1- \n)} (\a R)^2 + \cdots
&&
\end{flalign}
These show how $\ga(R)$ leaves the ``previously fixed points'' $\ga = \pm \n$. In each case we have a \emph{different} beta function, that can be expanded perturbatively near the respective points as
\begin{flalign}
\beta_\rm{IR}(\ga) &= 2 (\ga - \n) + \cdots	\label{betaIR}
\\
\beta_\rm{UV}(\ga) &= 2 (\ga + \n) + \cdots \label{betaUV}
\end{flalign}
Note that these functions are \emph{not} given by the limits of (\ref{beta1}) for $\ga \to \pm \n$.
In particular, while in (\ref{beta1}) the point $\ga = + \n$ was a IR stable fixed point (i.e.~with decreasing $\beta(\ga)$), here \emph{both} beta functions are UV stable (both are \emph{increasing}). 
This is expected: 
for finite $L$, one goes toward the point $\ga = + \n$ at \emph{large} $R$ but, despite the IR/UV nomenclature associated with the value of $L$, in Eqs.(\ref{gaRaIR})-(\ref{gaRaUV}) we now arrive at \emph{both} fixed points $\ga = \pm \n$ by taking a \emph{small} distance cutoff $R$. So both cases are UV limits. 

\bigskip

Choosing $L = L_\rm{IR}$ or $L_\rm{UV}$ corresponds to choosing one of the (sets of) integration constants $A_k$ or $B_k$ to vanish; in the IR, $B_k = 0$, and, in the UV, $A_k = 0$.
In the strongly repulsive region of parameter space, normalizability of $\psi_k(x)$ \emph{enforces} $L = L_\rm{IR}$, recovering the asymptotic conformal symmetry. Similarly, the discussion of Sect.\ref{bound_c} shows that we can use SUSY to reestablish the asymptotic conformal symmetry in part of the weak-medium region where the Pöschl-Teller potential is supersymmetric, by setting $L = L_\rm{IR}$.
Or, from the opposite point of view, the presence of an anomalous scale $L > 0$ spontaneously breaks not only the asymptotic conformal symmetry, but also SUSY.

\subsubsection{The critical line $\n = 0$}

The two fixed points $\ga = \pm \n$ merge if $\n = 0$. The critical line in the $g_c$-$g_s$ plane where this happens is precisely the line $g_s = - \frac14$, which marks the boundary between the medium-weak and strongly attractive regions, cf.~Fig.\ref{Moduli_plane}. On this critical line, solution (\ref{sol_u}) is not valid; the correct solution of the Schrödinger equation is the one given in Eq.(\ref{psi_0_geral}) in Appendix \ref{appen}. Because of this, we cannot simply take the limit $\n \to 0$ in the formulas for the RG flow derived above, since they were all deduced from the solution (\ref{sol_u}).

The details of the renormalization and the computation of the running coupling in this case are given in Appendix \ref{appen}. Using the asymptotic solution (\ref{x_0_0}) with $k=0$, we can compute the running coupling 
\be	\label{gamma_0M}
\gamma(R) = - \frac{{\cal D}}{1 - {\cal D} \log(\a R)}
\ee
where ${\cal D}$ is a \emph{dimensionless} constant, that plays a paper similar to $L$, cf.~Eq.(\ref{DefcalD}).
Interestingly, now $\a R$ appears logarithmically in $\ga(R)$. The presence of the logarithmic branch in Eq.(\ref{EDO_u}) is precisely why this solution must be considerably separately.
The beta function $\beta = d \ga / d \log R$ reads
\be	\label{beta0}
\beta(\ga)  = - \ga^2 + \cdots
\ee
which agrees with Eq.(\ref{beta1}) in the limit $\n \to 0$. The UV and IR fixed points of the conformal beta function, $\ga = \pm \n$, here coalesce into one \emph{marginal} fixed point $\ga = 0$. This is a kind of Berezinskii-Kosterlitz-Thouless (BKT) phase transition \cite{Berezinsky:1972rfj,Berezinsky:1970fr}. 

The dots in Eq.(\ref{beta0}) hide terms that are perturbative in $\a R$, as in (\ref{beta1}). But because there is also a logarithmic contribution of $\a R$, here we cannot separate the effects of ``spontaneous versus explicit'' breaking of conformal symmetry, as done above. Instead, the scale $1/\a$ appears along with the parameter ${\cal D}$, which is related to the ratio of integration constants of the solution (\ref{psi_0_geral}). So, even without taking the $\a R$ corrections into account, we see that, as $R$ increases, the running coupling (\ref{gamma_0M}) diverges in a massive RG flow with mass scale
\be	\label{LEgnthsdsa}
1/L_0 = \a e^{- 1/{\cal D}} .
\ee
We could suppress this massive phase by suppressing the logarithmic solution near the origin, which amounts to choosing the corresponfing integration constants of the Schrödinger equation as ${\cal B}_k = 0$, which makes ${\cal D} = 0$; cf.~Eq.(\ref{renor_0}).
Then we would need to compute the perturbative powers of $\a R$ in (\ref{beta0}) to find the (explicit) breaking of conformal symmetry.

\subsubsection{Strongly attractive coupling and the phase $\theta$}	\label{SectRenorStrontehta}

For the strongly attractive regime, where $g_s<- \frac14$, and
$\nu=i|\nu|$, instead of the anomalous length scale $L$, renormalization introduces a \emph{phase} $\theta$. 
Here, the constants $A_k$ and $B_k $ are complex, and conjugates of one another, cf.~Eq.(\ref{x_<}). For $k > 0$, we can parameterize them in a convenient polar form in terms of real functions $C_k $ and $\zeta(k)$, 
\be	\label{A_B_theta}
A_k = \frac{k^{i|\nu|}}{2i} C_k  e^{i\zeta(k)} = B_k^* .
\ee
For $k = 0$ this parametrization is ill defined, and instead we choose
\be	\label{ABattrcreg}
A_0 = \frac{\a^{i|\nu|}}{2i} C_0  e^{i\theta} = B_0^* .
\ee
Then the asymptotic behavior (\ref{x_<}) reads
\bsub	\label{x_<_}
\begin{align}
\psi_k (x) &\approx C_k  \, x^{\frac12} \, \sin \big[ |\nu| \log (k x)+\zeta(k) \big] \\
\label{x_<_0}
\psi_0 (x) &\approx C_0  \, x^{\frac12} \, \sin \big[ |\nu| \log (\a x) + \theta \big].
\end{align}
\esub

The function ${\cal F}(R)$ can be computed from Eq.(\ref{Robin}) using $\psi_0(x)$ and $\psi_k(x)$. The renormalization framework identifies the two results, yielding
\be	\label{ren_g_forte}
\left( k / \a \right)^{2i|\nu|} e^{2i\zeta(k)} = e^{i\theta}, 
	\qquad 0 \leq \theta < 2\pi
\ee
which can also be written as
$|\nu| \log \left( k/\a \right) + \zeta(k) = \theta - n\pi,$ 
with $n \in {\bb N} .$
For bound states, this equation discretizes the wave vector $k$.
If the potential were exactly the conformal potential, then $\zeta$ would be independent of $k$, and we would have $k_n = k_0 e^{- n\pi / |\nu|}$ \cite{renor_orig,Renor_Inver_Squa,renor_Group_Limit_Cycle}.
For the Pöschl-Teller potential, the phase $\zeta(k)$ is determined in terms of the arbitrary renormalization parameter $\theta$, and fixed by the other boundary condition at $x \to \infty$; see \S\ref{sec:lig_theta}.

Computing ${\cal F}(R) = \psi'_0(x) / \psi_0(x)$ from the asymptotic solution (\ref{x_<_0}), we find the running coupling (\ref{gamma}) to be
\be	\label{F_k0_2}
\ga(R) = |\n| \cot \big[ |\n| \log(\a R) + \theta \big] .
\ee
Again, we have a logarithmic contribution from $\a R$, as in the $\n = 0$ case. Again, this stems from a logarithmic branch in the solution (\ref{x_<_}). The corresponding asymptotic beta function is, again,  as in Eq.(\ref{beta1}), but with $\nu = i | \nu|$,
\be	\label{beta2}
\beta(\ga) =  - \ga^2 - |\n|^2  + \cdots
\ee
Now there are no fixed points, not even at $R = 0$. 
The two points $\ga = \pm \n$, which merged in the BKT phase transition when $\n = 0$, now have completely disappeared, and conformal symmetry is necessarily broken. 
The RG flow is massive, with every value of $\ga(R)$ reached for a \emph{finite} range of $R$. There are two ends of the massive flow, with mass scales $\a e^{\theta / |\n|}$ and $\a e^{(\theta - \pi)/|\n|}$.
As in the $\n = 0$ case, we cannot separate the effects of the explicit and the spontaneous breaking of conformal symmetry, because $\a$ necessarily appears in the mass scales of the RG flow together with $\theta$, even if we neglect perturbative corrections to (\ref{beta2}).

\section{Spectra of the Pöschl-Teller Hamiltonian}	\label{SectSpectra}

The presence of the scale $L$ or the phase $\theta$ influences the spectrum of the Pöschl-Teller Hamiltonian --- for example, it could be expected that, as it breaks the asymptotic conformal symmetry, $L$ may produce bound states confined to the region very near the singularity. Every aspect of the spectrum can be read from the S-matrix
\be
S \equiv e^{2i \delta(k)} ,
\ee
defined by the phase shift $\delta(k)$ such that the asymptotic solution for $kx \gg 1$ is
\be	\label{sin}
\begin{aligned}
\psi_k(x)
	&\sim e^{i\delta(k)} \sin\left[k(x-\Delta x)+\delta(k)\right]
\\
	&= \sin\left[k(x-\Delta x)\right] 
			+ e^{i\delta(k)} e^{ik(x-\Delta x)} \sin\delta(k) .
\end{aligned}
\ee
We will now compute the S-matrix of the Pöschl-Teller potential for any value of the renormalization scale $L$, including the UV and IR fixed points (\ref{UVIRL}) of the asymptotic RG flow.
As it was to be expected, the spectrum depends sensibly on the region of the $g_s$-$g_c$ plane, besides the value of $L$ or $\theta$.

Expanding the hypergeometric functions in (\ref{sol_u}) for $kx \gg 1$  (cf.~e.g.~\cite[\S15]{NIST:DLMF}),
\be	\label{x_inf}
\begin{aligned}
&\psi_k(x) \approx
\\
&\quad
	\Bigg[
	\frac{A_k}{\alpha^{\frac12 + \n}} \
	\frac{
	\Gamma(1+\nu)\Gamma (-ik / \a )
	}{
	\Gamma (\frac{1 + \n + \m - ik/\a}{2})
	\Gamma (\frac{1 + \n - \m - ik/\a}{2})
	}
	+ 
	\frac{B_k }{\a^{\frac12 - \n}}
	\frac{
	\Gamma(1-\nu) \Gamma (- ik / \a )
		}{
	\Gamma (\frac{1 - \n + \m - ik/\a}{2})
	\Gamma (\frac{1 - \n - \m - ik/\a}{2} )
		}
	\Bigg] \frac{e^{-ik x}}{2^{-\frac{ik}{\alpha}}}
	\\
&\quad
	+
	\Bigg[\frac{A_k}{\a^{ \frac12 + \n}}
	\frac{
	\Gamma(1+\nu) \Gamma (ik / \a )
		}{
	\Gamma (\frac{1 + \n + \m + ik/\a}{2})
	\Gamma (\frac{1 + \n - \m + ik/\a}{2})
		}
	+ \frac{B_k }{\a^{\frac12 - \n}} 
	\frac{\Gamma(1-\nu) \Gamma(ik / \a)
		}{
	\Gamma (\frac{1 - \n + \m + ik/\a}{2} )
	\Gamma (\frac{1 - \n - \m + ik/\a}{2})
		}
	\Bigg] \frac{e^{ik x}}{2^{\frac{ik}{\a}}} .
\quad
\end{aligned}
\ee
In the medium-weak and in the strongly repulsive regime, we can phrase the boundary conditions in the language of the RG flow to rewrite $S$ in terms of the renormalization scale $L$.
Comparison of (\ref{sin}) and (\ref{x_inf}) shows that $\Delta x \equiv \a^{-1} \log 2$, and the S-matrix is
\be	\label{delta3}
\begin{aligned}
S^\vare_L(k) = e^{2i\dexp } \
	&\Bigg[
	\frac{
	\Gamma(1+\nu)
	}{
	\Gamma (\frac{1 + \n + \m + ik/\a}{2}  )
	\Gamma (\frac{1 + \n - \m + ik/\a}{2}  )
	}
+\vare (\alpha L)^{2\nu}
	\frac{
	\Gamma(1-\nu)
	}{
	\Gamma (\frac{1 - \n + \m + ik/\a}{2})
	\Gamma (\frac{1 - \n - \m + ik/\a}{2})
	}
	\Bigg]
\\	
\times
	&\Bigg[
	\frac{
	\Gamma(1+\nu)
	}{
	\Gamma (\frac{1 + \n + \m - i k / \a}{2}  )
	\Gamma (\frac{1 + \n - \m - ik/\a}{2}  )
	}
	+\vare (\alpha L )^{2\nu}
	\frac{
	\Gamma(1-\nu)
	}{
	\Gamma (\frac{1 - \n + \m - ik/\a}{2} )
	\Gamma (\frac{1 - \n - \m - ik/\a}{2} )
	}
	\Bigg]^{-1} ,
\end{aligned}
\ee
where
\be	\label{delta_exp}
\exp 2i \dexp
	\equiv
 	- \frac{\Gamma (ik/\a )}{\Gamma (- ik/\a )} .
\ee
This phase $\dexp$, which does not depend on the parameters $\m$, $\n$ is a universal feature of S-matrices of potentials that behave as ${\cal V}(x) \sim e^{-\a x}$ for $x \to \infty$. Although it contributes to the phase shift $\delta(k)$, it will not contribute to the analysis of the spectrum to be done below.
Since $L$ is independent of $k$, Eq.(\ref{delta3}) gives the explicit dependence of the S-matrix on the wave number $k$.
It is a general expression that reduces to special cases when we choose specific boundary conditions, i.e.~choose $L$. If the potential is strongly attractive, instead of $L$ we must use the phase $\theta$ defined in Eq.(\ref{ren_g_forte}), 
\be	\label{delta4}
\begin{aligned}
S_{\theta}(k)
	= e^{2i\dexp } \
	&\Bigg[
	\frac{\Gamma(1+i|\nu|)
		}{
	\Gamma (\frac{1 + i |\n| + \m + ik/\a}{2})
	\Gamma (\frac{1 + i|\n| - \m + ik/\a}{2})
	}
	-
	\frac{
	e^{-i\theta} \ 
	\Gamma(1-i|\nu|)
	}{
	\Gamma (\frac{1 - i |\n| + \m + ik/\a}{2} )
	\Gamma (\frac{1 - i|\n| - \m + ik/\a}{2})
	} \Bigg]
\\	
	\times
	&\Bigg[
	\frac{
	\Gamma(1+i|\nu|)
	}{
	\Gamma (\frac{1 + i|\n| + \m - ik/\a}{2})
	\Gamma (\frac{1 + i|\n| - \m - ik/\a}{2})
	}
	-
	\frac{
	e^{-i\theta} \ 
	\Gamma(1-i|\nu|)
	}{
	\Gamma (\frac{1 - i|\n| + \m -ik/\a}{2} )
	\Gamma (\frac{1 - i|\n| - \m - ik/\a}{2})
	}
	\Bigg]^{-1}.
\end{aligned}
\ee

The S-matrices above describe scattering off the Pöschl-Teller potential in all ranges of the parameters $g_c$-$g_s$. 
The potential always has a continuous spectrum for $E>0$, that is, for real $k>0$, since ${\cal V}(x) \to 0$ for large $x$.
For the continuum of states, the phase shift $\delta(k) = \frac1{2i} \log S(k)$ is a real number, as can be seen most easily from the first equality in (\ref{sin}), given that $\psi_k(x)$ is real.
Meanwhile, for some ranges of $g_s$ and $g_c$, the potential also develops metastable, bound or anti-bound states; these can also be discovered by looking at the analytic properties of the S-matrices.

Bound states have negative energy,
$E = \hbar^2 k^2 / 2m < 0$, hence 
\be	\label{kBound}
k = i \kappa , \qquad \kappa > 0 ,
\ee
is imaginary.
By definition the wave-function vanishes at infinity, which means that the first, divergent term $\sim e^{-ikx}$ in the r.h.s.~of Eq.(\ref{x_inf}), must be set to zero. This amounts to setting to zero the square brackets that appear in the denominator of the S-matrix (\ref{delta3}), that is, bound states are poles of $S(\kappa)$, excepted the poles of $e^{2i\dexp }$.

We can also have different kinds of metastable states. These are solutions whose stationary part behave as an outgoing wave at infinity, $\psi(x) \sim e^{ikx}$, hence they are poles of the S-matrix, but we allow the wave number to have an imaginary part,
\be	\label{QNM}
k = k_r - i \ka, 
\qquad 
k_r > 0
\qquad \ka > 0 ,
\ee
which leads to a decay of the amplitude.
More precisely, the stationary wave-function $\psi_k(x) \sim e^{ikx}$ is not square-integrable, but the time-dependence of the total wave function is given by 
\be	\label{ExpEt}
\exp \left( - \frac{iEt}{\hbar} \right) = 
	\exp \left( - \frac{i \hbar (k_r^2 - \ka^2) t}{2m} \right) 
	\
	\exp \left( - \frac{\hbar k_r \ka}{m} t \right) ,
\ee
where we have briefly reintroduced $\hbar$ and $m$ for reference.
So, if $\ka > 0$, the wave function decays exponentially with time, and the state is said to be metastable, with a lifetime 
\be	\label{metsblelifet}
\tau=\frac{2m}{\hbar k_r\ka}.
\ee
For $\ka > 0$, if $\ka$ is small, the metastable state has a large lifetime, and produces a `resonance' peak in the scattering cross-section. Poles with $k_r = 0$ and $\ka > 0$ are called `anti-bound states', as they correspond to the bound states (\ref{kBound}), but with negative wave number $\kappa = - \ka < 0$.
Anti-bound states are not normalizable (since $\psi \sim e^{\ka x}$ diverges), but the real exponential in (\ref{ExpEt}) suppresses the total wave function, producing a discrete energy spectrum.
If there are poles of the S-matrix with $\ka < 0$ and $k_r \neq 0$, the real exponential diverges and the system is unstable. 

\emph{For definiteness, in the following we will reserve the expression `metastable states' for $k_r > 0$ and $\ka > 0$, and `anti-bound states' for $k_r = 0$ and $\ka >0$.}

The generalized Pöschl-Teller potential appears as an effective potential for fluctuations of pure $d$-dimensional de Sitter spacetime, where $x$ corresponds to the tortoise radius of the static (causal) patch with center at $x = 0$ and cosmological horizon at $x = \infty$ \cite{QNM_dS,Lopez-Ortega:2006aal}. Fluctuations are said to be `quasinormal modes' (QNMs) if they are purely \emph{outgoing} waves at $x = \infty$ and vanish at $x = 0$, and this gives the standard definition \cite{cardona_2017_PT,Lopez-Ortega:2006aal,QNM_BH_Julio,QNM_dS} of `QNM boundary conditions' for the generalized Pöschl-Teller potential (\ref{V_PT}).
Fluctuations of pure de Sitter must have $\psi(0) = 0$ because the values of $g_s$ that appear there are, almost always, in the strongly repulsive region of parameter space. 
Meanwhile, purely outgoing waves at $x = \infty$ are the fundamental characterization of QNMs as the dissipative fluctuations of a system: in pure or asymptotically de Sitter spacetimes, this corresponds to the fluctuations exiting the cosmological horizon \cite{Lopez-Ortega:2006aal,QNM_dS,Zhidenko:2003wq,Cardoso:2003sw}; without the cosmological horizon, e.g.~in Schwarzschild black holes, the purely outgoing waves are placed at spatial infinity.%
	\footnote{In the usual situation with black holes, the Pöschl-Teller potential has $g_s = 0$ and is defined on the whole real line. Then the QNM are defined by a purely outgoing wave at $x = \infty$ and a purely ingoing wave at $x = - \infty$, as the latter is the position of the black hole event horizon, see e.g.~\cite{Konoplya:2011qq,Berti:2009kk}.}

The condition $\psi(0) = 0$ corresponds, in our language, to putting $L = 0$, and is forced upon us in part of the $g_c$-$g_s$ plane; but in the remaining regions the renormalization procedure has given us a way of defining well-posed boundary conditions with finite (or infinite) $L$. 
Then we have the Robin boundary condition (\ref{Robin}) for $\psi(x)$ at the cutoff $x = R$, and an unphysical function $\psi^<(x)$ inside the regularized region $0 < x < R$. 
(Note that nothing prevents us to set $\psi^<(0) = 0$.)
Meanwhile, outgoing waves at infinity define the metastable and anti-bound states.
Thus the metastable and anti-bound states we will study in this section, over the entire $g_c$-$g_s$ plane, could be thought of, somewhat loosely, as ``quasinormal modes'', in the sense of being dissipative fluctuations --- even if the boundary conditions at the origin that we will use are not necessarily standard.

\subsection{Spectra for $L = 0$}	\label{SectSpectL0}

\begin{figure}[t] 
\centering
\includegraphics[scale=0.52]{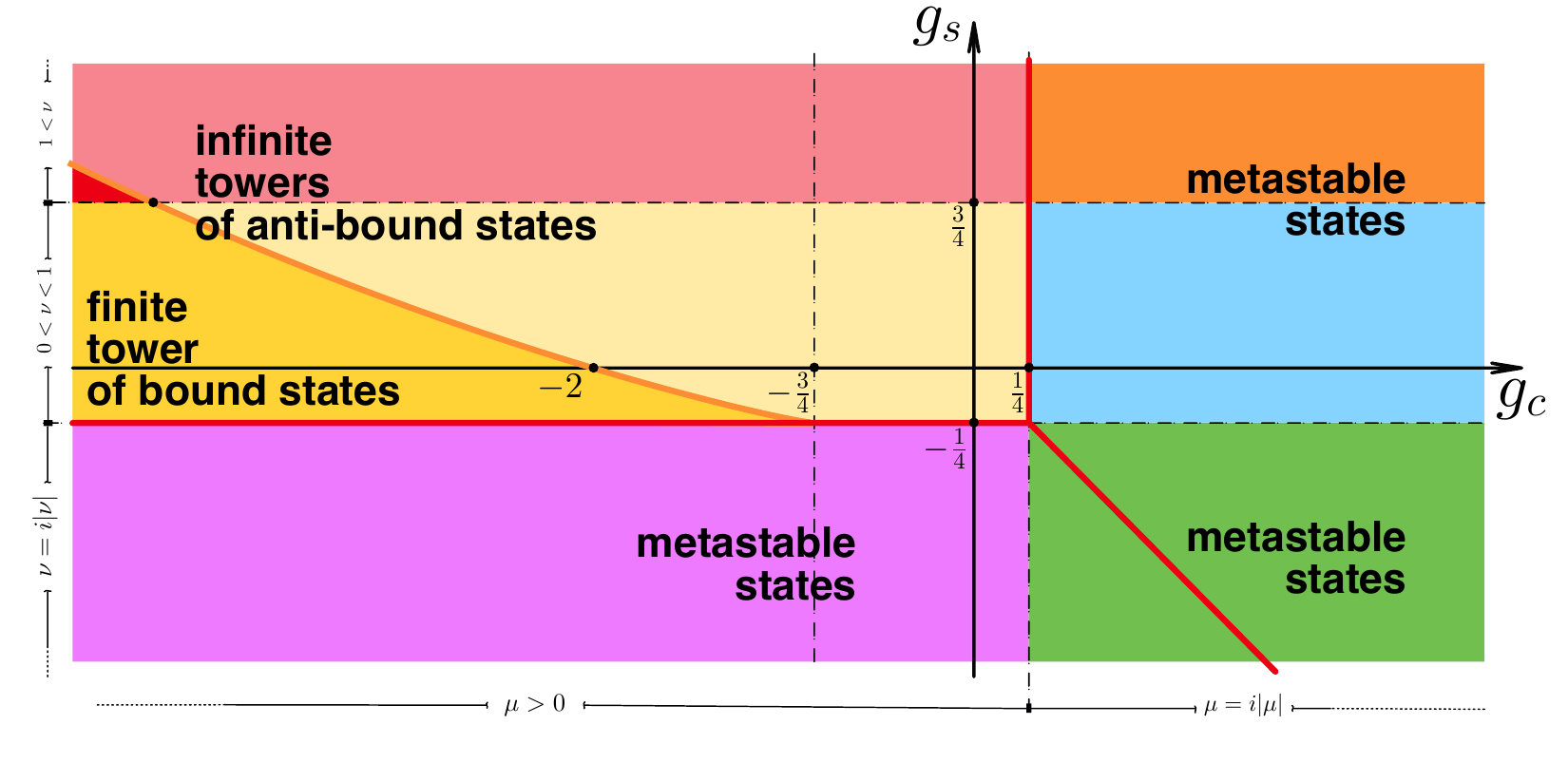}
\caption{
``Phase diagram'' for the spectra with $L = 0$. 
A finite tower of bound states, (\ref{IRboundsat}), exists only inside the darker wedge inside the red and yellow regions. Infinite towers of anti-bound states, (\ref{UV_anti}), exist in the entire (dark and light) yellow and red patches. A tower of anti-bound states, (\ref{RedLineAntib}), exists on the red line. In the orange, blue, green and purple regions, we have the metastable states (\ref{MetsbleIR}).
}
\label{Moduli_IR}
\end{figure}

Let us consider the spectrum for ``IR boundary conditions'', fixing the anomalous scale to $L = 0$.
This choice of $L$ is \emph{allowed} in every region of the $g_c$-$g_s$ plane, although it is only \emph{necessary} in the strongly attractive region. (If we require that the potential is supersymmetric, it becomes necessary also in part of the medium-weak region, as noted at the end of \S\ref{SextSpontBreakExpl}.)
A ``phase diagram'' showing the different regions of the $g_c$-$g_s$ plane with their different spectra is shown in Fig.\ref{Moduli_IR}.

The S-matrix (\ref{delta3}) simplifies considerably to
\be	\label{delta2}
S_\rm{IR} (k)
	= e^{2i\dexp } \
	\frac{
	\Gamma (\frac{1 + \n + \m - ik/\a}{2} )
	\Gamma (\frac{1 + \n - \m - ik/\a}{2} )
	}{
	\Gamma (\frac{1 + \n + \m + ik/\a}{2} )
	\Gamma (\frac{1 + \n - \m + ik/\a}{2} )
	}.
\ee
Apart from $\bar\delta(k)$, the poles of $S_\rm{IR}(k)$ are at the poles of the pair of Gamma functions in the numerator. 
We can see by inspection that there are no instabilities. 
Bound states have $k = i \kappa$, with $\kappa > 0$, so the Gamma functions only have poles if both $\m,\n \in \bb R$. This means that we are in the yellow and red patches of Fig.\ref{Moduli_IR}. The argument of the second Gamma function is a negative integer for
\bsub\label{IRboundsat}
\be\label{kappa1}
\kappa_n = \a (\m - \n - 1 - 2n  ) ,
\ee
where $n \in \bb N$.
Since $\kappa > 0$, there is a \emph{finite} number $n_\rm{max}$ of bound states, such that 
\be	\label{E_des}
0 \leq 2 n \leq 2 n_\rm{max} \leq \mu-\nu-1. 
\ee
\esub
In turn, (\ref{E_des}) means that bound states exist, i.e.~$n_\rm{max} \geq 0$,
only if 
$\mu \geq \nu+1.$
This condition defines a wedge on the $g_c$-$g_s$ plane, whose border is a curve drawn in dark orange in Fig.\ref{Moduli_IR}; we only have bound states inside the darker yellow, and darker red patches.

\subsubsection{Quasinormal modes}		\label{SectQNML0}

As mentioned above, for $L = 0$, since $\psi(0) = 0$, the metastable and anti-bound states correspond to the standard definition of quasinormal modes of the generalized Pöschl-Teller potential \cite{cardona_2017_PT,Lopez-Ortega:2006aal,QNM_BH_Julio,QNM_dS}.
That is, when the $g_s$ and $g_c$ have the appropriate values, the set of metastable and anti-bound states found from (\ref{delta2}) correspond to the QNM of pure de Sitter spacetime. We will borrow this terminology, but nevertheless be general about the value of the parameters $g_s$ and $g_c$. 

We are looking for poles of the S-matrix (\ref{delta2}) with $k = k_r - i \ka$, where $\ka > 0$ and $k_r \geq 0$. In general, one of the Gamma functions in the numerator we will have poles if
\be	\label{QNMkgen}
k = k_r - i \ka = - i \a \big( \n \pm \m + 1 + 2n \big) ,
\ee
with $n \in \bb N$.
In different regions of parameter space, $\m$ and/or $\n$ become real or imaginary, either contributing to the oscillating part $k_r$ or to the damping part $\ka$. 
For some different choices of $\m,\n$, Eq.(\ref{QNMkgen}) reduces to known formulae for QNM frequencies of the Pöschl-Teller potential \cite{Lopez-Ortega:2006aal,QNM_BH_Julio,QNM_dS}.

\begin{figure}[t] 
\centering
\includegraphics[scale=0.6]{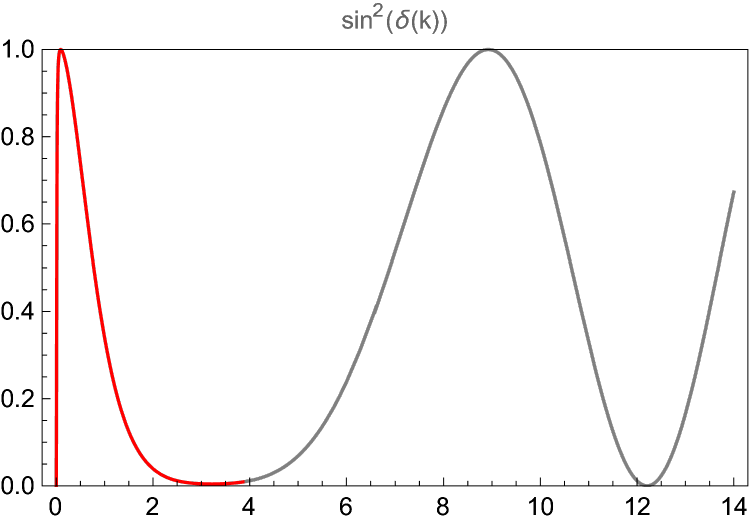}
\caption{Resonance at low energy due to an anti-bound state for $L = 0$; here we chose
$\n = 5.59$ and $\m = 6.60$.}
\label{fig:resson_UV_anti}
\end{figure}

In the red and yellow patches 
(both the light and darker regions)
of Fig.\ref{Moduli_IR}, where $\m,\n \in \bb R$, there are no metastable states with $k_r > 0$, only anti-bound states with $k_r = 0$; in fact, there are two series of anti-bound modes,
\bsub	\label{UV_anti}
\begin{align}
	\label{UV_anti_+}
\ka_+ &= + \m + \n + 1 + 2n_+ , 
	&& n_+ = 0 , 1 , 2, 3, \dots
\\
	\label{UV_anti_-}
\ka_- &= - \m + \n + 1 + 2n_- , 
	&& n_- = n_\rm{min}, \ n_\rm{min}+1, \ n_\rm{min}+2, \dots
\end{align}
\esub
In the second series, $n_\rm{min}$ is the smallest integer to satisfy the condition 
$n_\rm{min} > \m - \n - 1$. 
Note that the ``missing'' integers $0,1,\dots, n_\rm{min}-1$ are such that $\ka_ - < 0$; these are just a relabeling of the bound states (\ref{kappa1}).
There may be resonances in the series (\ref{UV_anti_-}), that happen at small energies if $\m \gtrsim \n +1$. We illustrate the presence of one such resonance in Fig.\ref{fig:resson_UV_anti}, with a plot of the scattered wave $\sin^2 \delta(k)$ with a red peak at low $k$.

There are metastable states with $k_r > 0$ in every region of the $g_c$-$g_s$ plane outside the red and yellow patches of Fig.\ref{Moduli_IR}. In each of the colored regions of Fig.\ref{Moduli_IR} the parameters $\m$ and $\n$ contribute to the damped factor $\ka$ differently:
\be	\label{MetsbleIR}
\frac{k}{\a} = \frac{k_r - i \ka}{\a}
	=
	\left\{
	\begin{aligned}
	&\pm  |\m| - i (1 + |\n| + 2n) \quad&\text{for $g_c > \tfrac14$ and $g_s > - \tfrac14$}
	\\
	& + |\n| - i (1 \pm |\m| + 2n) \quad&\text{for $g_c < \tfrac14$ and $g_s < - \tfrac14$}
	\\
	&  \pm |\m| + |\n| - i(1 + 2n) \quad&\text{for $g_c > \tfrac14$ and $g_s < - \tfrac14$}
	\end{aligned}
	\right.
\ee
with $n = 0,1,2,3,\dots$
These combinations of $g_s$ and $g_c$ correspond to the orange, purple and green patches of Fig.\ref{Moduli_IR}, respectively.
We see that there is still another series of anti-bound states with $k_r = 0$, on the red line traced in Fig.\ref{Moduli_IR}.
The line has one branch at $\m = 0$ separating the yellow/red and orange/blue patches, another branch the boundary between the yellow and purple patches, $\n = 0$, and goes into the green patch where $|\m| = |\n|$ with
\be	\label{RedLineAntib}
k = - i\ka = - i \a (1+2n) .
\ee
These anti-bound states are limits of metastable states in the various regions when $k_r \to 0$.
The most stable states are those with smallest ``overtone number'', $n = 0$. The lifetimes measured in units of $k_r$, are given by Eq.(\ref{metsblelifet}) as 
$\tau_0 k_r = 2m /\hbar \ka \sim 1 / (1 + |\n|)$. So even for $\n = 0$, the lifetimes are not parametrically large, and the metastable states (\ref{MetsbleIR}) do not produce resonances.

\subsection{Spectra for finite $L$ or $\theta$}	\label{SectSpectLthe}

\begin{figure}[t] 
\centering
\includegraphics[scale=0.45]{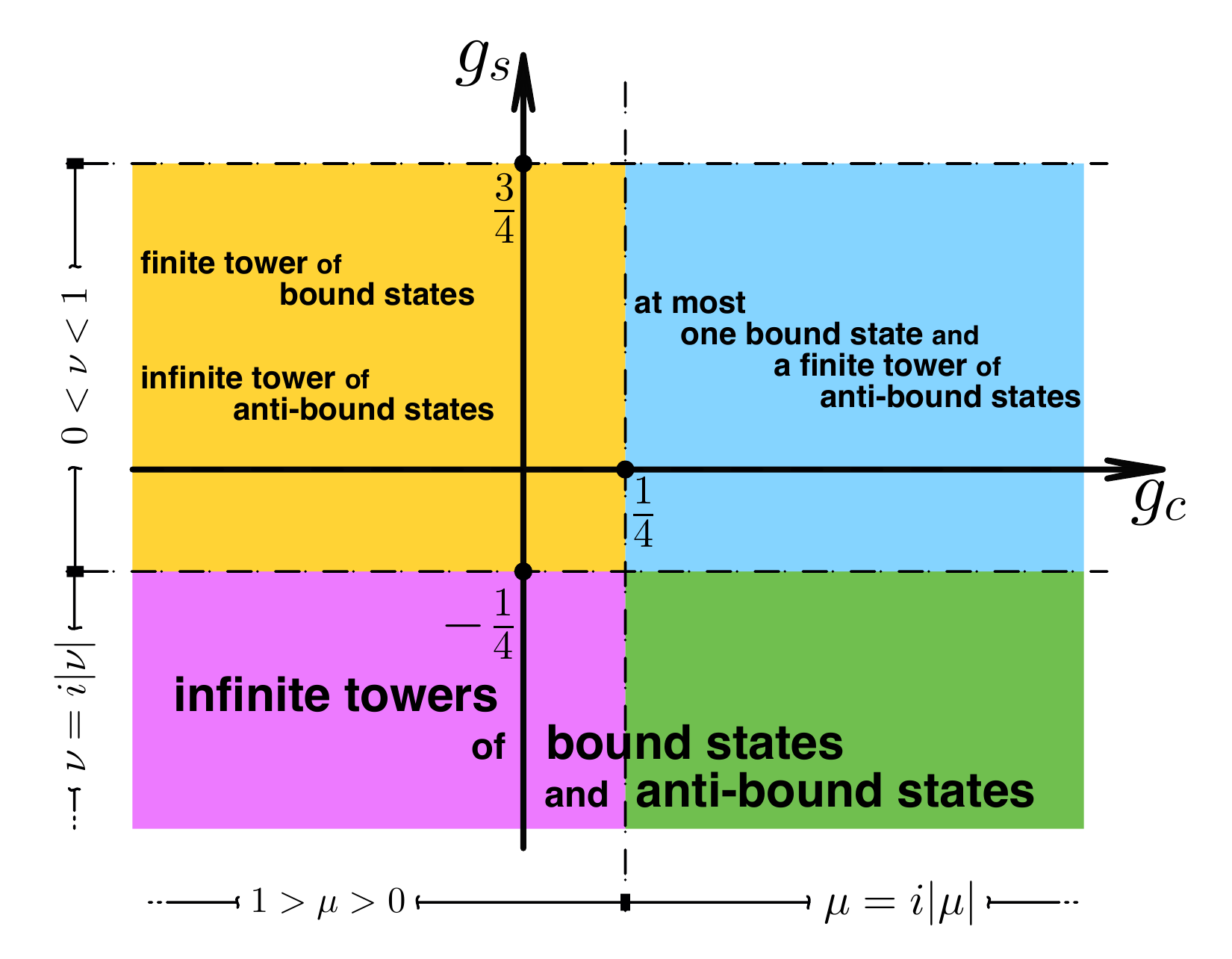}
\caption{%
``Phase diagram'' for the spectra with finite $L$ and $\theta$.
}
\label{Moduli_L_theta}
\end{figure}

In the regions of parameter space shown in Fig.\ref{Moduli_L_theta}, we are not forced to set  $L$ or the phase $\theta$ to any specific value. The spontaneous breaking of asymptotic conformal symmetry affects the spectrum; we find new bound and anti-bound states in regions of parameter space where there is none if $L$ or $\theta$ vanish.

\subsubsection{The medium-weak region}

Inspection of (\ref{delta3}) shows that we have poles of $S_L^\vare(k)$ when the dividing square bracket has a simple zero,
\be	\label{nonFgs}
- \vare (\alpha L)^{2\n}
	=
	\frac{\Gamma(1+\n)}{\Gamma(1-\n)} \
	\frac{
	\Gamma (\frac{1 - \n + \m - i k/\a}{2} )
	\Gamma (\frac{1 - \n - \m - i k/\a}{2} )
		}{
	\Gamma (\frac{1 + \n + \m - i k / \a}{2} )
	\Gamma (\frac{1 + \n - \m - i k/\a}{2} )
	} .
\ee
The l.h.s.~is a real number. If $k = k_r - i \ka$ with a non-vanishing $k_r > 0$, the function in the r.h.s.~is complex; hence \emph{there are no metastable states with $k_r > 0$ in the medium-weak region.}
We then search for bound/anti-bound states, making $k = i \kappa$ in Eq.(\ref{nonFgs}), 
 \be	\label{F}
- \vare (\alpha L)^{2\n}
	=
	\frac{\Gamma(1+\n)}{\Gamma(1-\n)} \
	\frac{
	\Gamma (\frac{1 - \n + \m + \kappa/\a}{2} )
	\Gamma (\frac{1 - \n - \m + \kappa/\a}{2} )
		}{
	\Gamma (\frac{1 + \n + \m +\kappa / \a}{2} )
	\Gamma (\frac{1 + \n - \m + \kappa/\a}{2} )
	}
\equiv {\scr F}(\kappa/\a;\n,\m).
\ee
if, given $L$, we find a $\kappa > 0$ that solves this equation, we get a bound state. If we find a solution for some $\ka = - \kappa > 0$, we get an anti-bound state.

\begin{figure}[t] 
\subfigure[
Black: plot of ${\scr F} (x;\n,\m)$, Eq.(\ref{F}), as a function of $x = \kappa/\a$, in the yellow patch of Fig.\ref{Moduli_L_theta}, where $\m,\n \in \bb R$.
Red: chosen value of $\vare (\a L )^\n$.
There is a finite tower of bound states (black dots) for $\kappa > 0$, and an infinite tower of anti-bound states (red dots) for $\kappa < 0$.
Here $\n = 0.7$ and $\m = 2.73$.
]
{
\centering
\includegraphics[scale=0.54]{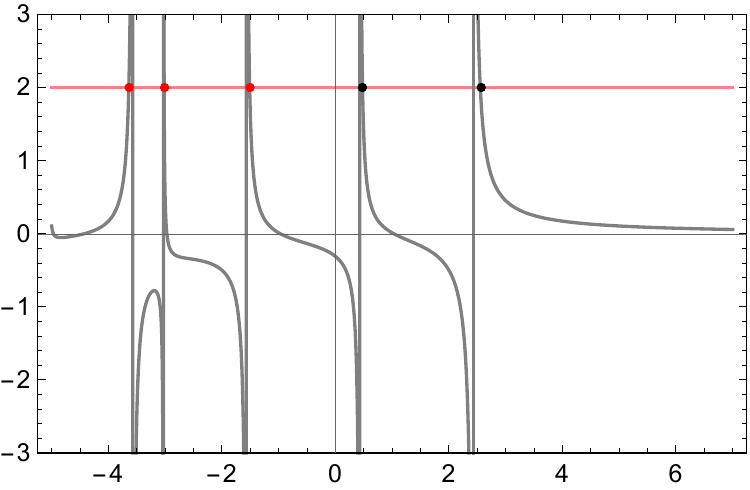}
\label{fig:lig_intermed_blue_patch}
}
\qquad
\subfigure[
Black: plot of ${\scr F} (x;\n,\m)$, Eq.(\ref{F}), as a function of $x = \kappa/\a$, in the blue patch of Fig.\ref{Moduli_L_theta}, where $\n \in \bb R$, $\m = i |\m|$.
Red: chosen values of $(\a L )^\n$.
If $L$ is large, there are no bound nor anti-bound states.
If $(\a L )^\n < {\scr F}(0;\n;|\m|)$, there is one bound state (black dot), and a finite number of anti-bound states (red dots).
Here $\n = 0.7$ and $\m = i 0.8$.
]
{
\centering
\includegraphics[scale=0.54]{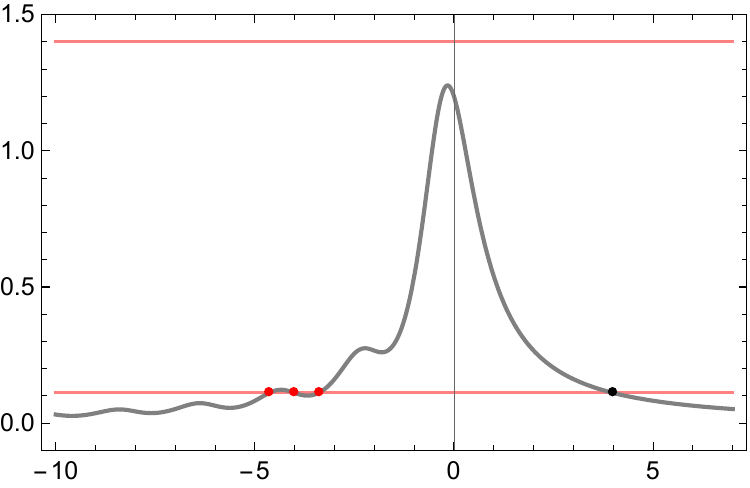}
\label{fig:lig_intermed}
}
\caption{}
\end{figure}

In the yellow patch, both $\m,\n \in \bb R$, and $0 < \n < 1$. One of the two gamma functions in the numerator will have poles at
$\kappa/\a = -1 + \n \mp \m - 2 n ,$
with
$n = 0,1,2,3,4,\dots$
This gives a finite set of poles of ${\scr F}(\kappa/\a)$ for $\kappa > 0$, for which the maximum value of $n$ is the largest integer that satisfies
\be
0 < n_\rm{max} \leq \tfrac12 (1 - \n \mp \m ).
\ee
We also have an infinite number of poles of ${\scr F}(\kappa/\a)$ for $\kappa = - \ka < 0$.
The function ${\scr F}(\kappa/a)$ is monotonic between its poles if $\kappa > 0$. This means that, for any choice of $\vare = \pm 1$, and any given $L$, there will be $n_\rm{max}$ solutions of Eq.(\ref{F}) with $\kappa > 0$, hence \emph{there are $n_\rm{max}$ bound states}; and there will be infinite solutions with $\ka = - \kappa > 0$, hence \emph{there are infinite anti-bound states}. A typical profile of ${\scr F}$ is plotted in Fig.\ref{fig:lig_intermed_blue_patch}.

In the blue patch of parameter space shown in Fig.\ref{Moduli_L_theta}, we still have $0 < \n < 1$, but now $\mu = i |\mu|$. 
Then the function ${\scr F}(\kappa/\a;\n,|\m|)$ is \emph{real} because the Gamma functions appear in products of the type $\Gamma(z) \Gamma(\bar z) = |\Gamma(z)|^2$.
Hence the second fraction in (\ref{F}) is never negative, neither is the first fraction, because $0 < \nu < 1$, and we conclude that ${\scr F}(\kappa/\a;\n,|\m|) \geq 0$.
As a result, Eq.(\ref{F}) does not have a solution if the l.h.s.~is negative, that is $S^+_L(\kappa)$ has no poles, and \emph{there are no bound nor anti-bound states} in this case. 
If the r.h.s.~of Eq.(\ref{F}) is positive, we may have solutions or not. 
In Fig.\ref{fig:lig_intermed} we plot the typical behavior of ${\scr F}$ as a function of $\kappa / \a \geq 0$. It is a decreasing, monotonic function, which goes to zero at infinity, and has its largest value at $\kappa = 0$.
As a consequence, Eq.(\ref{F}) has only one solution if 
\be
0 < (\a L)^{2\n} < {\scr F}(0; \n; |\m|) ,
\ee
and no solution if $(\a L)^{2\n} > {\scr F}(0; \n; |\m|)$.
In short, $S^-_L(\kappa)$ can have at most one pole, hence \emph{at most one bound state, depending on the scale parameter $L$.}
As for anti-bound states, for $\kappa = - \ka < 0$ the function ${\scr F}(\kappa/\a;\n,|\m|)$ oscillates and goes to zero for $\kappa \to - \infty$. So, depending on the value of $L$, we get \emph{at most a finite number of anti-bound states; if $L$ is large enough, we have none.}
A typical profile of ${\scr F}$ is plotted in Fig.\ref{fig:lig_intermed}.

\subsubsection{The strongly attractive region; ``Efimov effect''}\label{sec:lig_theta}

In the strongly attractive region $g_s<- \frac14$, the S-matrix is given by Eq.(\ref{delta4}), in terms of the arbitrary phase $\theta$.
Now $\n = i | \n|$ is an imaginary number in the positive imaginary axis.
We are in the green and purple regions of Fig.\ref{Moduli_L_theta}.

Let us first search for bound states.
Set $k = i \kappa$, with $\kappa > 0$.
Once more, we search for zeros of the whole denominator of (\ref{delta4}), i.e.~solution of 
\be	\label{f_g}
e^{i\theta}
	=
	\frac{\Gamma(1-i|\nu|) }{ \Gamma(1+i|\nu|)}
	\frac{
	\Gamma (\frac{1+ i|\n| + \m + \kappa/\a}{2})
	\Gamma (\frac{1 + i|\n| - \m + \kappa/\a}{2})
		}{
	\Gamma (\frac{1 - i|\n| + \m + \kappa/\a}{2})
	\Gamma (\frac{1 - i|\n| - \m + \kappa/\a}{2})
		}
\equiv 
\exp i f (\kappa/\a; |\n|,\m ).
\ee
By fixing the relation between $\theta$ and $k$, this equation also fixes the phase $\zeta(k)$ in Eq.(\ref{ren_g_forte}) and completes the description of the eigenstates.

\begin{figure}[t] 
\subfigure[Plot of
 $f(x;|\nu|,\mu) \in[-\pi,\pi)$, Eq.(\ref{f_g}), as a function of $x = \kappa/\a$; vertical lines mark discontinuities between $\pm \pi$.
Red: a chosen value of $\theta$.
Black dots: bound states. 
Here $|\nu|=3$, $\mu=0.5$.
]{
\centering
\includegraphics[scale=0.5]{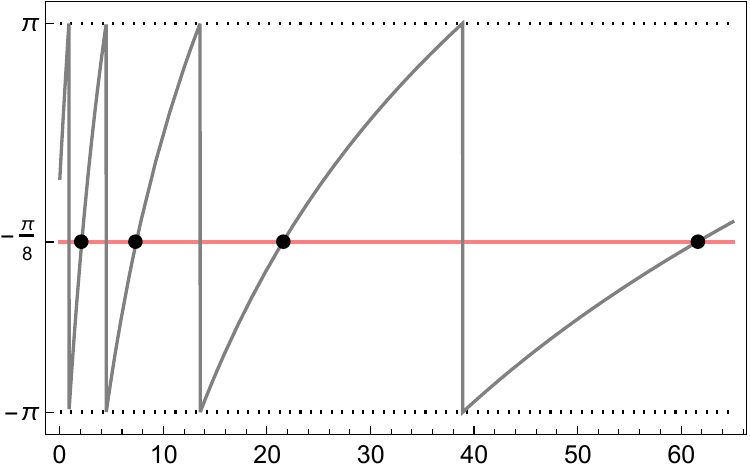} 
\label{fig:estados_lig}
}
\qquad
\centering
\subfigure[Blue: plot of $\bar f( x; | \n| , \m)$, Eq.(\ref{bar_f_g}), as a function of $x = \ka / \a$, with $k_r = 0$; vertical lines are the discontinuities between $\pm \pi$.
Red: a chosen value of $\theta$. Black dots: anti-bound states.]
{
\includegraphics[scale=0.5]{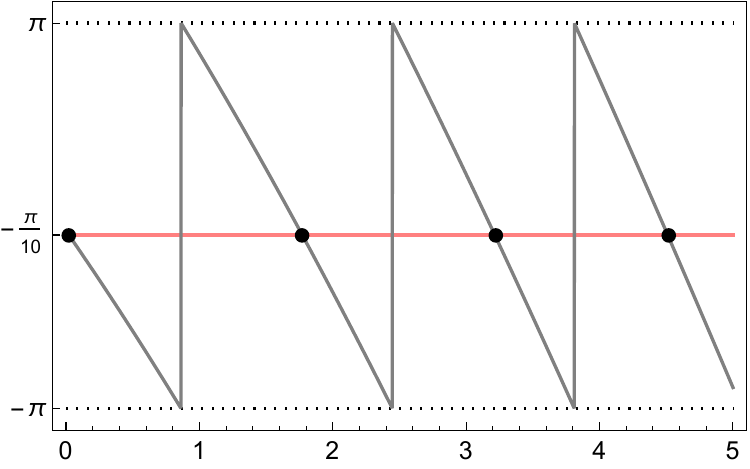}
\label{fig:anti_lig_theta}
}
\caption{}
\end{figure}

The bound states are those $\kappa$ which solve Eq.(\ref{f_g}), i.e.~for which $\theta = f(\kappa/\a; |\n|, \m)$.
The analysis of this equation is more conveniently done graphically. In Fig.\ref{fig:estados_lig} we plot $f(x; |\n|, \m)$ as a function of $x = \kappa/\a$, for representative values of $|\n|$ and $\m$. We see that there is an infinite number of bound states given by the solutions $\kappa_n$. There is no defined value for the lowest energy, i.e.~no maximum value of $\kappa_n$. Moreover, the lower the energy $E_n \sim - \kappa_n^2$, the larger is the gap $\Delta E_{n,n+1}$ between consecutive bound states.
The numerical values of the first five solutions $\kappa_n$ of Eq.(\ref{f_g}), for the parameters in $|\n| = 3$, $\m = 0.5$ used in the example of Fig.\ref{fig:estados_lig}, are
\begin{equation*}
\frac{\kappa_1}{\alpha}=2.1178, \quad
\frac{\kappa_2}{\alpha}=7.3824, \quad
\frac{\kappa_3}{\alpha}=21.5399, \quad
\frac{\kappa_4}{\alpha}=61.5593, \quad
\frac{\kappa_5}{\alpha}=175.4851.
\end{equation*}
The ratios between consecutive levels,
 \begin{equation*}
 \frac{\kappa_2}{\kappa_1}=3.4859, 
\quad
\frac{\kappa_3}{\kappa_2}=2.9177, 
\quad
\frac{\kappa_4}{\kappa_3}=2.85792, 
\quad \frac{\kappa_5}{\kappa_4}=2.85067,
\end{equation*}
approach the value $e^{\pi / |\n|} \approx 2.8497$ for $|\n| = 3$.
That is, as we increase $n$,
\be	\label{ratikap}
\kappa_{n+1}/\kappa_n \approx e^{\pi /|\nu|}, 
\qquad n \gg 1.
\ee

Recall that here we are in the strongly attractive regime; the bound states are essentially locked inside the pit near the origin, where  ${\cal V} \approx g_s / x^2$. 
This is why the behavior of bound states that we find here is similar to the Efimov effect for the conformal potential \cite{Efimov}. However, here the bound states do not accumulate near the threshold of zero energy: for small $\kappa$ the wave functions with $E \approx 0$ spread farther from the origin, and feel the difference between the Pöschl-Teller and the conformal potentials. 
The example above corresponds to a point with $\m > 0$ (purple region in Fig.\ref{Moduli_L_theta}), but the results are analogous for imaginary $\m$ (green region in Fig.\ref{Moduli_L_theta}).

\bigskip

\begin{figure}[t] 
\centering
\includegraphics[scale=0.6]{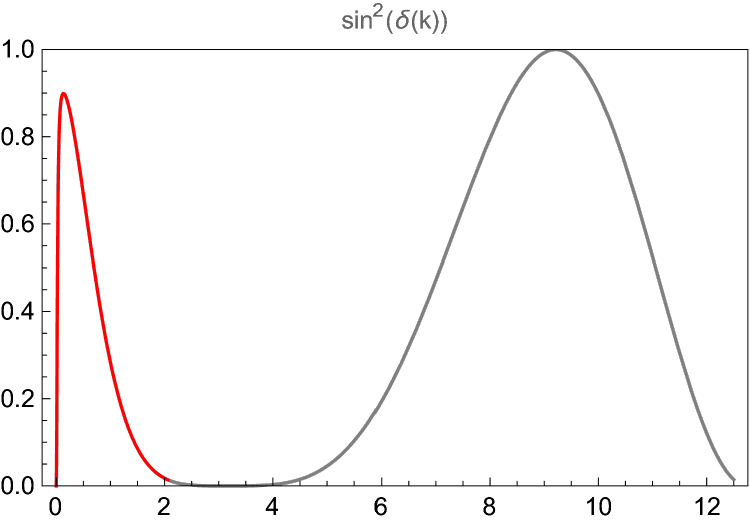} 
\caption{Scattered wave amplitude:  at low energies, the lowest anti-bound state (\ref{menor_lig}) creates the resonance marked in red.
Here $|\nu|=3.00$, $\mu=1.72$ and $\theta=-\frac{\pi}{10}$.}
\label{fig:sin_res_g_forte}
\end{figure}

We now search for metastable states.
For a given (real) $\theta$, the poles of the S-matrix (\ref{delta4}) with $k = k_r - i\ka$ 
are the solutions of
\be	\label{bar_f_g}
e^{i\theta}
	=
	\frac{\Gamma(1-i |\nu|)}{\Gamma(1+i |\nu|)}
	\frac{
	\Gamma (\frac{1 + i |\n| + \m}{2} - \frac{\ka + i k_r}{2\a} )
	\Gamma (\frac{1 + i|\n| - \m }{2}  - \frac{\ka + ik_r}{2\a} )
		}{
	\Gamma (\frac{1 - i|\n| + \m}{2}  - \frac{\ka + ik_r}{2\a} )
	\Gamma (\frac{1 - i|\n| - \m}{2} - \frac{\ka + ik_r}{2\a} )
	}
	\equiv \exp i \bar f(\ka/\a; |\n|,\m) .
\ee
For the equation to have a solution, we must have $\bar f \in \bb R$, which means that the combination of Gamma functions must be a phase, i.e.~its modulus squared must be one. This holds if and only if $k_r = 0$; in this case, the every Gamma function in the numerator of (\ref{bar_f_g}) is the complex conjugate of the corresponding Gamma function in the denominator below, hence the square modulus of the expression is one.
If $k_r \neq 0$, this matching of complex conjugate numbers is spoiled.
Since $k_r$ is forced to vanish, \emph{there are no metastable states in the strongly attractive region.}
Setting $k_r = 0$, we find \emph{an infinite series of anti-bound states}, with $\ka > 0$. Again, this is best seen graphically, and shown in Fig.\ref{fig:anti_lig_theta}.
We plot $\bar f(\ka/\a; | \n| , \m) \in [-\pi , + \pi]$. The anti-bound states are at the intersections with a horizontal line that marks a chosen value of $\theta$.
In our example, we chose 
$|\n| = 3.00$, $\m=1.72$ and $\theta = - \frac{\pi}{10}$; in this case, the first anti-bound state has
\be	\label{menor_lig}
\ka_\rm{min}/\alpha = 0.02179,
\ee
which is small enough to produce a resonance of the scattered wave amplitude $\sin^2 \delta(k)$ at the low energies, as shown in Fig.\ref{fig:sin_res_g_forte}.

\subsection{Spectra for $L=\infty$}	\label{SectUVspectrum}

\begin{figure}[t] 
\centering
\includegraphics[scale=0.4]{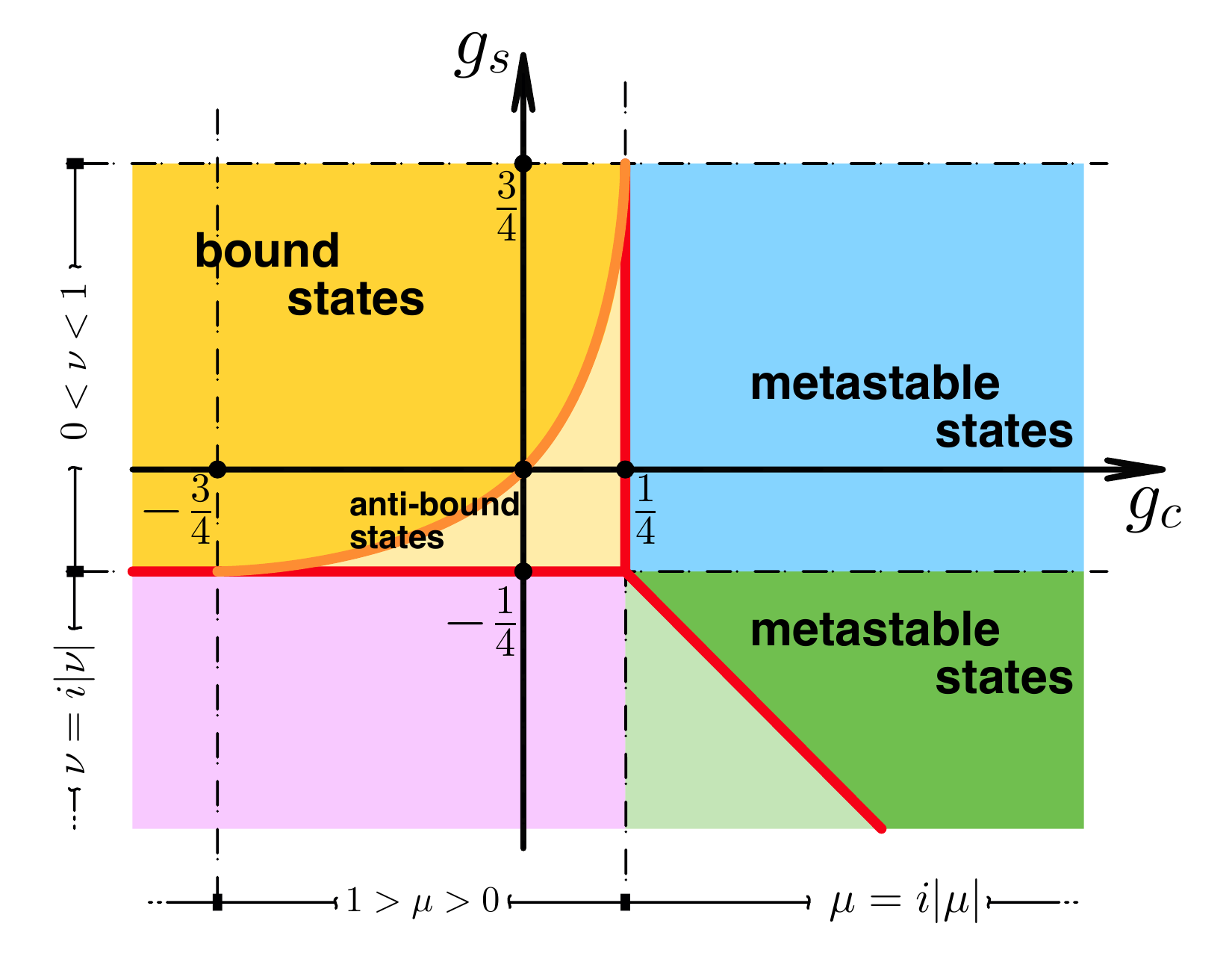}
\caption{``Phase diagram'' for the spectra with $L = \infty$. 
A finite tower of bound states exists in the dark yellow patch, (\ref{kappa1UV}).
Two infinite towers of anti-bound states exists in the entire (dark and light) yellow patch, (\ref{towerantigyll}).
Infinite towers of metastable states exist in the blue and dark green patches, Eqs.(\ref{k_qn_IR}) and (\ref{k_qn_IR_b}) respectively.
Metastable states degenerate into anti-bound states on the red lines, (\ref{antibounpurl}), (\ref{k_qn_IR_antd}) and (\ref{k_ab_IR_b}).
There are no metastable, bound nor anti-bound states in the light green and purple patches.
}
\label{Moduli_L_theta_UV}
\end{figure}

At the ``UV point'' of the RG flow, setting $L = \infty$, the S-matrix (\ref{delta3}) simplifies to
\be	\label{delta_UV}
S_\rm{UV} (k)
	= e^{2i\dexp } \
	\frac{
	\Gamma (\frac{1 -\n + \m - ik/\a}{2} )
	\Gamma (\frac{1 - \n - \m - ik/\a}{2})
	}{
	\Gamma (\frac{1 - \n + \m + ik/\a}{2})
	\Gamma (\frac{1 - \n - \m + ik/\a}{2} )
	}.
\ee
As in the case of finite $L$, we can only choose $L = \infty$ if we are not forced to choose $L = 0$, so the strongly repulsive region is forbidden. Thus we are, once again, restricted to the patches of the $g_c$-$g_s$ plane shown in Fig.\ref{Moduli_L_theta_UV}, which should be compared with the ``phase diagram'' in the same regions, shown in Fig.\ref{Moduli_L_theta}, obtained for finite $L$ (and $\theta$).

First let us consider the yellow patch, part of the medium-weak region, where $\m, \n \in \bb R$. Bound states are the poles of the Gamma functions in the numerator of (\ref{delta_UV}) with $k = i \kappa$, and $\kappa > 0$.
Since $0 < \n < 1$, only the second Gamma function has poles, yielding \emph{a finite tower of bound states}, with
\be	\label{kappa1UV}
\kappa_n = \a (\mu + \n - 1 - 2n ), 
\qquad 0 \leq n \leq n_\rm{max} .
\ee
and $n$ integer. 
As before in \S\ref{SectSpectL0}, the condition that $\kappa > 0$ means there is a maximum number of bound states $n_\rm{max}$, such that 
$0 \leq 2n \leq 2n_\rm{max} \leq \m + \n  - 1$.
Imposing that $n_\rm{max} \geq 0$ gives a restriction in the $g_c$-$g_s$ plane in terms of a curve $\m + \n = 1$, drawn in orange in Fig.\ref{Moduli_L_theta_UV}.
(Compare (\ref{kappa1UV}) to (\ref{IRboundsat}) for bound states in the same region in parameter space, but with $L = 0$, and the patches in Figs.\ref{Moduli_L_theta_UV} and \ref{Moduli_IR}.)

\begin{figure}[t] 
\centering
\includegraphics[scale=0.55]{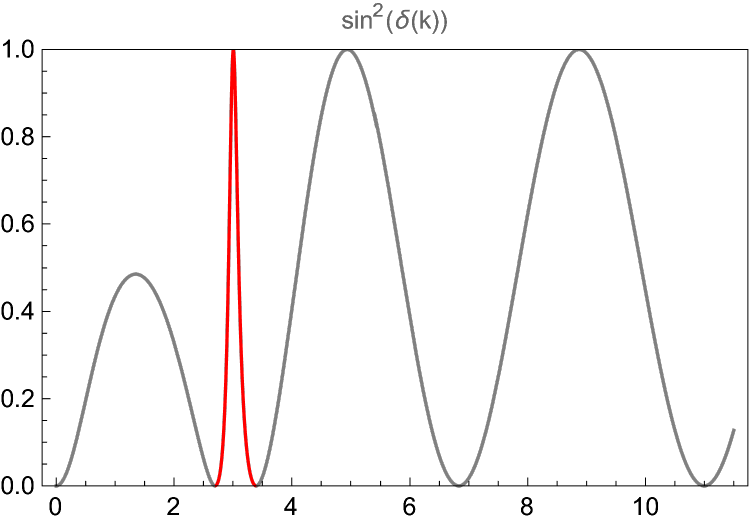} 
\caption{Plot of scattering amplitude with resonance around $k/\a = |\m|$ marked in red. Here $\n = 0.9$ and $|\m| = 3.0$.}
\label{fig:resson_QNM_IR}
\end{figure}

Inspection of (\ref{delta_UV}) shows that \emph{there are no bound states outside the yellow patch}. To find metastable and anti-bound states, we take $k = k_r - i \ka$, and search for poles of the Gamma functions in the numerator of (\ref{delta_UV}).
These happen when there are solutions of
\be	\label{ksoltexuv}
k = k_r - i \ka = i  \a (\n \mp \m)  - i \a (1 + 2n) ,
\qquad k_r \geq 0 , \quad \ka \geq 0,
\ee
with $n = 0,1,2,3,\dots$
The solutions, when they exist, depend on the region of parameter space.
First, let us consider again the yellow patch, where $\m,\n \in \bb R$. Clearly there are no metastable states with $k_r \neq 0$, and we only have two towers of anti-bound states with
\be	\label{towerantigyll}
k_\pm = - i \a (1 - \n \mp \m + 2n) .
\ee
If $\m + \n > 1$, the tower $k_-$ starts at $n_\rm{min} = \lceil \frac12 ( 1 -\n - \m) \rceil$; the missing states with $n < n_\rm{min}$ are the finite tower of bound states (\ref{kappa1UV}).

Now, take
 $\n = i |\n|$ and $\m > 0$, in the purple patch of Fig.\ref{Moduli_L_theta_UV}; the real part on the r.h.s.~of (\ref{ksoltexuv}) is $- |\n| < 0$, so there are no solutions for $k_r$, hence \emph{no anti-bound nor metastable states inside the purple region.}
However, when $\n = 0$, we have two towers of anti-bound states at the threshold of the yellow and purple regions, with
\be	\label{antibounpurl}
k_\pm = - i \a (\pm \m + 1 + 2n) .
\ee
If $\m > 1$, the tower $k_-$ starts from $n_\rm{min} = \lceil \frac12 ( 1 - \m) \rceil$. This is the limit of the tower (\ref{towerantigyll}); again, the missing anti-bound states are the bound states in (\ref{kappa1UV}).

In the blue patch of Fig.\ref{Moduli_L_theta_UV}, since $0 < \n < 1$ and $\m$ is imaginary, there are solutions of (\ref{ksoltexuv}) with the minus sign,
\be	\label{k_qn_IR}
k = k_r - i \ka =   \a |\m| - i \a  \big( 1-\n + 2n \big) 
\ee
with $n = 0,1,2,3,\dots$
Since $0 < \n < 1$ in the blue patch, the system is stable because $\ka > 0$.
Hence we have \emph{metastable states which, at the threshold of the blue region, where $\m = 0$, become a series of anti-bound states}
\be	\label{k_qn_IR_antd}
k = - i \a  \big( 1-\n + 2n \big) 
\ee
The first metastable (or anti-bound) mode has $\ka_0 = \a (1-\n)$, which becomes arbitrarily small if $\n \lesssim 1$, i.e.~near the threshold $g_s \approx \frac34$. In this case, the lifetime $\tau_0 \sim 1 / (1-\n)$ becomes arbitrarily large, creating a resonance, as illustrated in Fig.\ref{fig:resson_QNM_IR}.
Finally take $\n = i |\n|$ and $\m = i |\m|$, in the green patch of Fig.\ref{Moduli_L_theta_UV}. 
If $|\m| - |\n| > 0$, the solutions of (\ref{ksoltexuv}) are \emph{a tower of metastable states},
\be	\label{k_qn_IR_b}
k = k_r - i \ka = \a (|\m| - \n) - i \a (1+2n) ,
\qquad
|\m| - | \n| > 0 .
\ee
In parameter space, the condition $|\m| > | \n|$ gives a triangular region, the dark green patch in Fig.\ref{Moduli_L_theta_UV}.
On the border of this region given by the diagonal line
$|\m| = |\n|$, the metastable states (\ref{k_qn_IR_b}) become \emph{a tower of anti-bound states} with
\be	\label{k_ab_IR_b}
k = - i \a (1 + 2n)  , \qquad |\m| = |\n| .
\ee

\subsection{Completing the spectrum in the medium-weak region}\label{sec:reta}

In the medium-weak region of parameter space, we can extend the spectrum of the Pöschl-Teller potential over the singularity at $x = 0$, to functions defined on the entire real line. This is equivalent to a self-adjoint extension of the Hamiltonian, but here done under the interpretation of the renormalization group. 

To illustrate the procedure clearly, consider first  consider a special case.
The medium-weak region of parameter space, where the boundary conditions are \emph{not} fixed by normalizability at $x = 0$, contains the line
\be	\label{nonSingLine}
g_s = 0 \qquad \text{hence} \qquad \nu = \tfrac12 ,
\ee
and $g_c$ arbitrary. 
For $\nu=\frac12$, Eq.(\ref{Robin}) becomes
\be	\label{Robin2}
\psi'_{k}(R)-\frac{1}{R + \vare L}\psi_{k}(R)=0,	
\ee
which has a well-defined $R \to 0$ limit,
\be	\label{Robin0}
\psi'_{k}(0) - \frac{\vare}{L} \psi_{k}(0)=0, 
\ee
so the Robin boundary conditions can be transferred directly to the origin, with no need for a cutoff. Setting $L = 0$ gives the Dirichlet condition $\psi_k(0) = 0$, while $L = \infty$ gives the Neumann condition $\psi_k'(0) = 0$.

On the line (\ref{nonSingLine}), the Pöschl-Teller potential is not singular at the origin, it degenerates into
\be	\label{calV0}
\mathcal{V}_0(x)=\frac{\alpha^2 g_c}{\cosh^2(\alpha x)},
\ee
which has no singularity, and is well-defined for all $x \in \bb R$ in the entire real line.
The spectrum is well-known. For $g_c < 0$, we have a finite tower of bound states with $k_N = i \kappa_N$,
\be	\label{kappa_reta}
\kappa_N = \a \left( \m - \tfrac12 - N \right) > 0, 
\qquad N = 0,1,2,3,\dots
\ee
We thus have bound states with both odd and even $N$, corresponding to the two  well-defined parities of the eigenstates of the symmetric potential ${\cal V}_0(x) = {\cal V}_0(-x)$.
Making $\a \mapsto i\a$, this 

Now, if we take the limit $g_s \to 0$, hence $\n \to \frac12$, in the spectrum obtained in Eq.(\ref{kappa1}), we do find (\ref{kappa_reta}), but \emph{only the odd modes} 
$N = 2n+1$.
What happens is that, to get to (\ref{kappa1}), we had fixed $L = 0$, which amounts to imposing a Dirichlet condition at the origin, $\psi(0) = 0$, and this condition only holds for the odd-parity states of the symmetric potential ${\cal V}_0$.
The even-parity states of ${\cal V}_0$, with $N = 2n$ are those that satisfy the Neumann condition $\psi'(0) = 0$ instead. 
These even states can also be found in our renormalization scheme, but they are at found with $L = \infty$.
Indeed, setting $\n = \frac12$ in the bound states (\ref{kappa1UV}), we find $\kappa_n = \a (\mu - \tfrac12 -2n )$,
which are precisely the even modes, $N = 2n$, in (\ref{kappa_reta}).

To summarize: to find all the eigenstates of the degenerate potential (\ref{calV0}), \emph{we must start with two independent renormalization scales $L_\rm{IR}$ and $L_\rm{UV}$, then set $L_\rm{IR} = 0$ and $L_\rm{UV} = \infty$.}%
	\footnote{%
	A similar approach was used in \cite{Gango_book} to find the spectrum on the entire real line of the harmonic oscillator added of a term $\sim 1/x^2$. The authors impose a small cutoff to the superpotential, which generates a delta-function spike to the potential, providing an explicit representation of the regularization procedure.}

The procedure above can be used to extend the eigenfunctions of the Pöschl-Teller potential, not only in the degenerate case $g_s = 0$, but on all of the medium-weak region in parameter space. 
Bound states with odd modes $\kappa_n^-$ are given by the poles of $S_\rm{IR}(\kappa)$, and those with even modes $\kappa^+_n$ by the poles of $S_\rm{UV}(\kappa)$, which were all found in (\ref{kappa1}) and (\ref{kappa1UV}).
Thus the spectrum of bound states for medium-weak potentials extended to the entire real line is
\be
\kappa_n^\pm = \a (\m \pm \n - 2n - 1 ), \qquad n \in {\bb N} ,
\ee
with contributions from both $L = 0$ and $L= \infty$.

\section{Conclusion}	\label{SectConclusion}	

In this paper, we have used a renormalization procedure to obtain well-defined energy eigenfunctions of the generalized Pöschl-Teller potential on the entirety of its parameter plane $g_c$-$g_s$, including the regions where the singularity at the origin renders the boundary conditions non-trivial.
Renormalization introduces an anomalous length scale $L$ (or a phase $\theta$ in the strongly attractive regime) by dimensional transmutation, and this parameterizes the family of solutions of the Schrödinger equation. 

The renormalization group of the Pöschl-Teller potential has a complicated structure. The anomalous scale $L$ competes with the intrinsic length scale of the potential, $1/\a$, for the breaking of asymptotic conformal symmetry near the origin. Near the singularity the beta function becomes that of the conformal potential, with two ``fixed points'', one UV and another IR, which are, however, only approximate, since conformal symmetry eventually is always broken by $\a$. For finite $L$, including the effects of $\a$ in the beta function is difficult, even perturbatively, but we were able to compute them in the two cases where asymptotic conformal symmetry is \emph{not} explicitly broken, namely $L = 0$ and $L = \infty$, corresponding to the asymptotic beta function being fixed in the IR or the UV ``fixed points'', respectively. Although the asymptotic beta function is the same in both cases --- it vanishes --- the $\a$-corrections are different, a clear example of how the two-parameter RG flow is complicated.

We have studied how the beta function and the running coupling behave in different patches of the $g_s$-$g_c$ plane. When we cross the line from the medium-weak to the strongly attractive regions of parameter space, the two asymptotic fixed points merge, undergo a BKT phase transition, then disappear.
In the strongly repulsive region of parameter space, the scale $L$ must be set to vanish by the normalizability of the wave function. We have shown that in the patch of the medium-weak region where the Pöschl-Teller potential is supersymmetric, the anomalous scale spontaneously breaks SUSY, along with the asymptotic conformal symmetry. 
From the opposite point of view, we can think of this as saying that SUSY can be used to fix $L = 0$ and prevent spontaneous breaking of asymptotic conformal symmetry, a phenomenon analogous to the one described by some of the present authors in \cite{Lima:2019xzg} for the exactly conformal potential. 

From the wave functions obtained by the renormalization procedure, we computed the S-matrix in all regions of the $g_c$-$g_s$ plane, and classified the stable, metastable and bound states.
The spectrum depends on the value of the anomalous dimension $L$ or the phase $\theta$, as was to be expected, and we produced ``phase diagrams'' showing the change of the spectra across the $g_c$-$g_s$ plane, for the qualitatively different choices of $L =0$, $L = \infty$, and $L$ or $\theta$ finite, Figs.\ref{Moduli_L_theta}, \ref{Moduli_IR} and \ref{Moduli_L_theta_UV}.
Looking at these ``phase diagrams'' we can track how bound, anti-bound and metastable states appear and disappear as one changes the parameters of the Pöschl-Teller potential, as well as the effect of spontaneous breaking of conformal symmetry induced by the anomalous dimension.
Finally, we discussed how the extension of the Hamiltonian to the entire real line requires information about the spectra at both ``fixed points'', $L = 0$ and $L = \infty$, simultaneously.

\bigskip

The explicit breaking of asymptotic conformal symmetry is a general feature of all potentials that have an inverse-square singularity, and it would be interesting to study our construction of the RG flow in more generality.
We would like, in particular, to study those potentials which, like the Pöschl-Teller, are supersymmetric and shape-invariant \cite{dutt1988}, including the class of potentials constructed in \cite{Odake2009,Odake2011,Odake2010,bougie2010}.
Note that all the shape-invariant potentials which are independent of $\hbar$ have been classified. Those which are restricted, by a singularity, to half the real line all diverge as $\sim g / x^2$, see \cite{dutt1988}. These potentials might have uses in the study of fluctuations of naked singularities \cite{Chirenti:2012fr,QNM_BH_Julio}.

We would also like to see whether the RG flow described here could have some holographic interpretation. Note that our procedure can be applied to the trigonometric version of the Pöschl-Teller Hamiltonian, which, in place of the hyperbolic functions, have a sine and a cosine. This potential appears in perturbations of AdS spacetime, and a thorough study of this Hamiltonian, including its self-adjoint extension has been done in \cite{Ishibashi:2004wx}. It would be interesting to contrast that analysis with our methods, considering that the Robin boundary conditions that we have found in the course of the renormalization procedure are the same that appear in \cite{Robin_1,Robin_2}.

From a different perspective, the trigonometric Pöschl-Teller potential describes a unidimensional Bravais lattice. In this case, when present, SUSY trivializes the lattice structure and does not allow energy bands \cite{Cooper:2001zd}. But in the regions of parameter space where there is no SUSY, renormalization should introduce a family with an infinite number of parameters. Perhaps these could be fixed to generate Bloch boundary conditions and non-trivial lattice structures.

\bigskip

We would like to consider the results we have obtained for the spectra in the contexts of several known applications of the Pöschl-Teller potential, such as quasinormal modes of black holes and, in particular, pure de Sitter spacetimes 
\cite{Ferrari:1984zz,Beyer:1998nu,
Berti:2009kk,
QNM_dS,Lopez-Ortega:2006aal,
Molina:2003ff,
Konoplya:2011qq,cardona_2017_PT,QNM_BH_Julio,Robin_1,Robin_2,Churilova:2021nnc,Jaramillo:2020tuu},
as well as fluctuations of relativistic kinks \cite{mendoncca20152,mendoncca20151,mendoncca2019,Zhong,Bazeia:2022yyv}.
As mentioned in Sect.\ref{SectSpectra}, the standard definition of quasinormal mode boundary conditions for the generalized Pöschl-Teller potential is that $\psi(x)$ is a purely outgoing wave at $x = \infty$ and $\psi(0) = 0$, see \cite{cardona_2017_PT,Lopez-Ortega:2006aal,QNM_BH_Julio,QNM_dS}.
It is the condition of outgoing waves at $x = \infty$ that characterize the QNMs as dissipative fluctuations. The condition at $x = 0$ is necessary in de Sitter spacetime because the corresponding potential is (almost always) strongly repulsive. In our general setting, where Robin boundary conditions are allowed in some regions of the $g_c$-$g_s$ plane, the anti-bound and metastable states found in Sect.\ref{SectSpectra} could perhaps be called ``quasinormal modes'' as well, despite the non-standard boundary conditions allowed at $x = 0$. 
We would like to explore whether this ``expanded definition'' of QNM does actually have interesting physics. For instance, in pure 3-dimensional de Sitter space, s-wave fluctuations lie outside the strongly repulsive region of the $g_c$-$g_s$ plane. Also, as recently discussed in \cite{QNM_BH_Julio}, self-adjoint extensions of the generalized Pöschl-Teller Hamiltonian have subtle effects on the mathematics of QNM, which would be interesting to consider under the light of the results of \S\ref{sec:reta}.

\subsection*{Acknowledgments}

AAL and UCdS would like to thank GM Sotkov for stimulating conversation; AAL also thanks BC Cunha and JP Calvalcante.
CFSP is funded by CAPES. AAL thanks the Bulgarian NSF for support.

\appendix

\section{$\a$-corrections to the running coupling}\label{ApednSwhofom}

In this appendix, we show what happens to the RG flow if we consider a perturbative expansion in powers of $\a R$. 
As stated in (\ref{alphaR1}), this is a small, dimensionless number. It measures the cutoff $R$ against the  Pöschl-Teller intrinsic length scale $1/\a$. Taking $\a R$ as small, but not zero, shows how this scale $1/\a$ affects the RG flow.

If we compute the running coupling $\ga(R)$ with $\psi_0$, from the solution (\ref{sol_u}), we can organize the result as
\be	\label{gaRa3}
\ga(R) 
	=
	\n
	\frac{
	1 - \vare (L/R)^{2\n}
	+ \sum_{n=1}^\infty \big[ N^0_n + \tilde N^0_n \, \vare (L/R)^{2\n} \big] (\a R)^{2n} 
	}{
	1 + \vare (L/R)^{2\n}
	+ \sum_{n=1}^\infty \big[ D^0_n + \tilde D^0_n \, \vare (L/R)^{2\n} \big] (\a R)^{2n} 
	}
\ee
where $\vare$ and $L$ are defined in (\ref{varedef}) and (\ref{ren_g_int0}),
and the coefficients
$N^0_n$, $\tilde N^0_n$, $D^0_n$, $\tilde D^0_n$ are dimensionless constants depending only on $\m$, $\n$.

Now let us compute the equivalent of the r.h.s.~of Eq.(\ref{gamma}) using $\psi_k(x)$ with $k > 0$ instead of $\psi_0(x)$.
Since $x$ and $k$ always appear in the dimensionless combinations $\a x$ and $k/\a$ in the solution (\ref{sol_u}), we find that
\be	\label{almsGrak}
- \frac{1}{2} +  R \, \frac{\psi_k'(R)}{\psi_k(R)} 
	=
	\n
	\frac{
	1 - \frac{A_k}{B_k} R^{2\n}
	+ \sum_{n=1}^\infty \big[ N_n(k) + \tilde N_n(k) \, \frac{A_k}{B_k} R^{2\n} \big] (\a R)^{2n} 
	}{
	1 + \frac{A_k}{B_k}  R^{2\n}
	+ \sum_{n=1}^\infty \big[ D_n(k) + \tilde D_n(k) \, \frac{A_k}{B_k} R^{2\n} \big] (\a R)^{2n} 
	}
\ee
where the series' coefficients are polynomials in $k/\a$, of the form
\be
\begin{aligned}
N_n(k) = N_n^0 + \sum_{j=1}^{2n} n_j (k/\a)^j ,
\quad
\tilde N_n(k) = \tilde N_n^0 + \sum_{j=1}^{2n} \tilde n_j (k/\a)^j ,
\\
D_n(k) = D^0_n + \sum_{j=1}^{2n} d_j (k/\a)^j ,
\quad
\tilde D_n(k) = \tilde D_n^0 + \sum_{j=1}^{2n} \tilde d_j (k/\a)^j ,
\end{aligned}
\ee
with $n_j$, $\tilde n_j$, $d_j$, $\tilde d_j$ being dimensionless constants that do not depend on $k$.
The highest power of $k$ in each term of the series in (\ref{almsGrak}) matches the corresponding power  of $R$.%
	\footnote{This can be seen by noting that $k/\a$ only appears in the arguments $a$ and $b$ of the hypergeometrics $_2F_1(a,b;c;z)$; the power of $k$ then matches the powers of $ab$ in the Pochhammer symbols.}
So we have at least one factor of $\a R$ for each factor of $k/\a$ and, when the condition (\ref{kR11}) for an effective description of the singularity is imposed, we get
\be	\label{almsGrakksm}
- \frac{1}{2} +  R \, \frac{\psi_k'(R)}{\psi_k(R)} 
	=
	\n
	\frac{
	1 - \frac{A_k}{B_k} R^{2\n}
	+ \sum_{n=1}^\infty \big[ N_n^0 + \tilde N_n^0 \, \frac{A_k}{B_k} R^{2\n} \big] (\a R)^{2n} 
	}{
	1 + \frac{A_k}{B_k}  R^{2\n}
	+ \sum_{n=1}^\infty \big[ D_n^0 + \tilde D_n^0 \, \frac{A_k}{B_k} R^{2\n} \big] (\a R)^{2n} 
	}
\quad \text{for} \quad kR \ll 1.	
\ee
Now, while $k R \ll 1$, Eq.(\ref{Robin}) implies that the r.h.s.~of Eqs.(\ref{almsGrakksm}) and (\ref{gamma}) must match. The only way for this to happen is if $A_k / B_k = \vare L^{2\n}$, as stated in Eq.(\ref{ren_g_int}).
More specifically, taking only zero order terms in $\a R$, we find
\be	\label{almsGrakMai}
- \frac{1}{2} +  R \, \frac{\psi_k'(R)}{\psi_k(R)} 
	=
	\n
	\frac{ 1 - \frac{B_k}{A_k} R^{-2\n}}{1 + \frac{A_k}{B_k}  R^{-2\n}}
\ee
which must match the running coupling (\ref{gaRa2}), and we arrive again at Eq.(\ref{ren_g_int}).

\section{The critical line $g_s = - \frac14$}	\label{appen}

On the line in parameter space defined by $g_s = - \frac14$, we have $\n = 0$, and the second solution of the hypergeometric equation appearing in (\ref{sol_u}) is not valid, due to a logarithmic branch of Eq.(\ref{EDO_u}) at the origin. 
A second solution can then be given by an hypergeometric centered at some other singularity, e.g.~at $u = 1$. Thus, the function
\be	\label{c=1_1}
[ \tanh(\a x)]^{\frac{1}{2}} [\cosh(\a x)]^{-i \frac{k}{\a}}
	\,{}_2F_1 \left[
	\frac{1 - \m + ik/\a}{2} ,
	\frac{1 + \m + ik/\a}{2} ;
	1+\frac{ik}{\alpha};
	\frac{1}{\cosh^{2}(\alpha x)}\right] 
\ee
is a solution of (\ref{EDO_u}) for $\n = 0$; see e.g.~\cite[\S15]{NIST:DLMF}. This is a complex function, whose complex conjugate is also a solution of (\ref{EDO_u}). Since we want a real solution, we combine both, and take the real part of (\ref{c=1_1}) as the second linearly independent solution of (\ref{EDO_u}) when $\n = 0$. Thus the general wave function in this case reads
\be	\label{psi_0_geral}
\begin{aligned}
&\psi_k (x)
	=
	\frac{[\tanh(\a x)]^{\frac{1}{2}} }{\a^{\frac12}}
	\Bigg[
	{\cal A}_k
	[\cosh(\a x)]^{-i \frac{k}{\a}}
	\,{}_2F_1 \left(
	\tfrac{1 - \m + ik/\a}{2} ,
	\tfrac{1+\m+ik/\a}{2}; \
	1; \ \tanh^2(\a x)
	\right)
\\	
&
+	\frac{{\cal B}_k}{2}
\Bigg(
	[ \cosh(\a x) ]^{- \frac{i k}{\a}}
	\tfrac{\Gamma (\frac{1-\m + ik/\a }{2} )
		\Gamma (\frac{1 + \m + ik/\a}{2})
		}{
		\Gamma(1+\frac{ik}{\a})}
		\,{}_2F_1 \left(
		\tfrac{1 - \m + ik/\a}{2} ,
		\tfrac{1 + \m + ik/\a}{2} ; 
		1 + \tfrac{ik}{\a};
		\tfrac1{\cosh^{2} (\a x)}
		\right)
\\
&\qquad\quad
+ 
	[ \cosh(\a x) ]^{\frac{i k}{\a}}
	\tfrac{\Gamma (\frac{1-\m - ik/\a }{2} )
		\Gamma (\frac{1 + \m - ik/\a}{2})
		}{
		\Gamma(1-\frac{ik}{\a})}
		\,{}_2F_1 \left(
		\tfrac{1 - \m - ik/\a}{2} ,
		\tfrac{1 + \m - ik/\a}{2} ; 
		1 - \tfrac{ik}{\a};
		\tfrac1{\cosh^{2} (\a x)}
		\right)
\Bigg)
\Bigg]
\end{aligned}
\ee
The factors of Gamma functions were chosen such that, expanding the hypergeometrics near the origin we find
\be	\label{x_0_0}
\psi_k(x) \approx {\cal C}_k \, x^{1/2} \left[1-\frac{\mathcal{B}_k}{\mathcal{C}_k} \log(\a x)\right] ,
\ee
where
\be		\label{C}
{\cal C}_k
	\equiv 
	{\cal A}_k
		- {\cal B}_k
		\Bigg[2 \ga 
		+ \rm{Re}\Big[
		\psi^{(0)} \left(\tfrac{1-\m+ik/\a}{2} \right)
		+\psi^{(0)} \left(\tfrac{1+\m+ik/\a}{2} \right)
		\Big]
		\Bigg]
\ee
with $\ga$ the Euler-Mascheroni constant and $\psi^{(0)}(z)$ the digamma function \cite[\S5]{NIST:DLMF}.
The asymptotic expansion at infinity is a combination of ingoing and outgoing waves,
\be	\label{assint_0}
\begin{aligned}
\psi_k(x) \approx
	\frac1{\a^{\frac12}}
	&\Bigg[
	\frac{
	\Gamma\left(\frac{ik}{\alpha}\right)
	{\cal A}_k
		}{
		\Gamma (\frac{1+\m+ik/\a}{2} )
		\Gamma (\frac{1-\m+ik/\a}{2} )
		}
		+
		\frac{
		\Gamma (\frac{1-\m-ik/\a}{2} )
		\Gamma (\frac{1+\m-ik/\a}{2} )
			}{
		\Gamma(1-\frac{ik}{\a})
		}
		\frac{{\cal B}_k}{2}
		\Bigg]
		\frac{e^{ik x}}{2^{\frac{2ik}{\a}}}
\\
&
+ \text{Complex Conjugate}
\end{aligned}
\ee

\subsection*{Renormalization}

Using the asymptotic solution (\ref{x_0_0}) with $k=0$, we can compute ${\cal F}(R)$, cf.~Eq.(\ref{Robin}),
\be	\label{AppclF}
R{\cal F}(R) = \frac{1}{2} - \frac{{\cal D}}{ 1- {\cal D} \log(\a R)}, 
\ee
with 
\be	\label{DefcalD}
{\cal D} = {\cal B}_0 / {\cal C}_0 = {\cal B}_k / {\cal C}_k
\ee
independent of $k$.
Using Eq.(\ref{C}) we can write this as a relation between ${\cal A}_k$ and ${\cal B}_k$, 
\be	\label{renor_0}
\frac{{\cal A}_k}{{\cal B}_k}
	=
	\frac12 
	\Big[ 
	\psi^{(0)}\left(\tfrac{1-\m+ik/\a}{2} \right)
	+\psi^{(0)}\left(\tfrac{1+\m+ik/\a}{2}\right)+
	\text{C.C.}
	\Big]
	+\frac{1}{{\cal D}} + 2\gamma.
\ee

\subsection*{Spectrum at the fixed point}

Choosing ${\cal B}_k = 0$ is necessary when  $g_c<1/4$, i.e.~$\mu$ is real. In this case, Eq.(\ref{assint_0}) simplifies. 
In terms of the phase shift, we have the S-matrix $S(k ) = e^{2i\delta(k)}$ given by
\be	\label{S_0}
S_0(k ) 
	=
	- 
	\frac{
	\Gamma (\frac{ik}{\alpha} )
	\Gamma (\frac{1}{2}+\frac{\mu}{2}-\frac{ik}{2\alpha} ) 
	\Gamma (\frac{1}{2}-\frac{\mu}{2}-\frac{ik}{2\alpha} )
	}{
	\Gamma (-\frac{ik}{\alpha} )
	\Gamma (\frac{1}{2}+\frac{\mu}{2}+\frac{ik}{2\alpha} )
	\Gamma (\frac{1}{2}-\frac{\mu}{2}+\frac{ik}{2\alpha} )
	}.
\ee
This result is valid even for imaginary $\m$ ($g_c > \frac14$). From the S-matrix we find the bound, anti-bound and metastable states to be, respectively,
\begin{align}
\kappa_{n} &= \a  (\mu-2n-1 ), 
	&& \quad \ \, n=0,1,2,3, \dots
\\
\kappa^\rm{AB}_{n_{\pm}} &= \a  (2n_{\pm} \pm \m +1 ), 
	&& \left[ \begin{aligned} &n_{+}=0,1,2,3, \dots \\ &n_{-}=n^\rm{min}_{-}, \ n^\rm{min}_{-} + 1 ,\dots \end{aligned} \right.
\\
k_{n}^\rm{MS} &= \a  [ |\m| - i(2n+1)] , 
	&& \quad \ \, n=0,1,2,3, \dots	\label{QNM_0_0}
\end{align}
where $n^\rm{min}_-$ is the first integer to satisfy the inequality $2 n^\rm{min}_- > \m - 1$. Bound states exist for $g_c<- \frac34$ ($\mu>1$); metastable states for $g_c > \frac14$ ($\m = i |\m|$); anti-bound states for $g_c < \frac14$ ($\m > 0$). The poles of metastable states cannot get arbitrarily close to the real axis, and do not create resonances. When $\m \approx 1$, $g_c \approx - \frac34$, we have a strong resonance at low energies, created by the bound state if $\m > 1$ or by the anti-bound state if $\m < 1$. These results correspond to the limit $\n \to 0$ of the spectra computed at the two fixed points $L_\rm{IR}$ and $L_\rm{UV}$ in the main text. The bound and anti-bound states correspond to the IR fixed point, Eqs. (\ref{kappa1}), (\ref{UV_anti_+}) and (\ref{UV_anti_-}), and the metastable states are given by the UV fixed point, Eq.(\ref{k_qn_IR}).

\subsection*{Spectra outside the fixed point}

\begin{figure}[t] 
\centering
\includegraphics[scale=0.7]{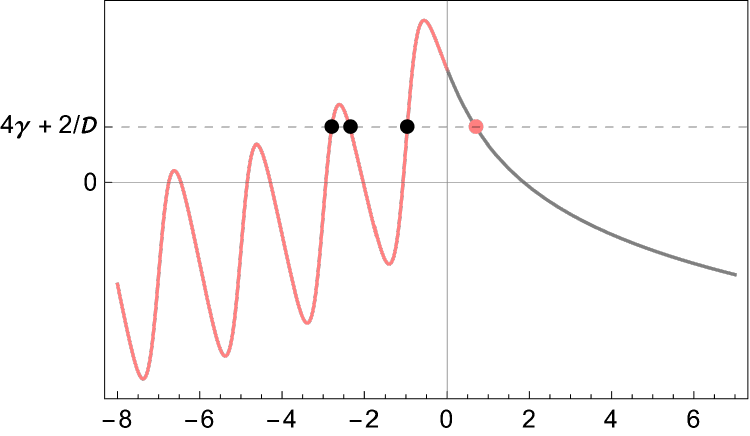}
\caption{Plot of ${\scr F}_0(x; i |\mu|)$ as a function of $x = \kappa/\a$. 
The horizontal line is a given value of $4\ga + 2 / {\cal D}$. Anti-bound states exist at the intersections when $\kappa < 0$, and bound states at the intersection when $\kappa > 0$.
}
\label{lig_anti_0_2}
\end{figure}

To be outside the fixed point, we must have $g_c>1/4$, hence $\m = i |\m|$.
The asymptotic solution (\ref{assint_0}) gives the S-matrix
\be	\label{S_0_L}
\begin{aligned}
S_{L_0}(k) = S_0(k)
	&
	\left[
	\frac{2 {\cal A}(k)}{{\cal B}(k)} 
	+ \frac{i\alpha}{k}
	\frac{
	\left|\Gamma (\frac{1- i|\m| - ik/\a}{2} )\right|^2
	\left|\Gamma (\frac{1 + i|\m| - ik/\a}{2} )\right|^2
		}{
	\left|\Gamma (\frac{ik}{\alpha} )\right|^2
	}
	\right]
\\
&\times
	\left[
	\frac{2 {\cal A}(k)}{{\cal B}(k)} - \frac{i \alpha}{k}
	\frac{
	\left|\Gamma (\frac{1- i|\m| - ik/\a}{2} )\right|^2
	\left|\Gamma (\frac{1 + i|\m| - ik/\a}{2} )\right|^2
		}{
	\left|\Gamma (\frac{ik}{\alpha} )\right|^2
	}
	\right]^{-1}
\end{aligned}
\ee
Here we used that $\Gamma(1+x)=x\Gamma(x)$, and $2{\cal A}(k)/{\cal B}(k)$ is given by (\ref{renor_0}) in terms of the length scale (\ref{LEgnthsdsa}) defined by ${\cal D}$. 
We have factorized the S-matrix at the fixed point $S(k)$, given by Eq.(\ref{S_0}). Since the poles of $S_0(k)$ are not poles of the new factor in $S_{L_0}(k)$, the metastable states (\ref{QNM_0_0}) are poles of $S_{L_0}(k)$, so they still exist outside the fixed point. The new poles are the solutions of 
\be	\label{L_0_polos}
\frac{2}{{\cal D}} + 4\ga = {\scr F}_0(k/\alpha;i|\mu|) ,
\ee
where 
\begin{equation*}
\begin{aligned}
{\scr F}_0(k/\alpha;i|\mu|)
	&\equiv
	\frac{\alpha}{\kappa}
	\frac{
	\Gamma (\frac{1 - i|\m| - ik/\a}{2} )
	\Gamma (\frac{1 - i|\m| + ik/\a}{2} )
	\Gamma (\frac{1 + i|\m| - ik/\a}{2} )
	\Gamma (\frac{1 + i|\m| + ik/\a}{2} )
	}{
	\Gamma (-\frac{ik}{\alpha} )
	\Gamma (\frac{ik}{\alpha} )
	}
	\\
	&-
	\Bigg[
	\psi^{(0)} \left( \tfrac{1 - \m + ik/\a}{2} \right)
	+\psi^{(0)}\left(\tfrac{1 + \m + ik/\a}{2} \right)
	\psi^{(0)} \left( \tfrac{1 - \m - ik/\a}{2} \right)
	+\psi^{(0)}\left(\tfrac{1 + \m - ik/\a}{2} \right)
	\Bigg].
\end{aligned}
\end{equation*}
Depending on the value of ${\cal D}$, we may have no bound nor anti-bound states; up to two anti-bound states and no bound state; or a finite number of anti-bound states and one bound state. In Fig.\ref{lig_anti_0_2} we have a typical curve of ${\scr F}_0(k/\alpha;i|\mu|)$; the bound and anti-bound states are the points where the curve touches the horizontal line $2/{\cal D} + 4\ga$. In the example shown, there are three anti-bound states (black dots), and one bound state (red dot). 
Finally, by direct inspection it is easy to see that there are no metastable states, since the right-hand-side of Eq.(\ref{L_0_polos}) is complex for $k=k_r-i\kappa$.

\bibliographystyle{utphys}
\bibliography{PT_generalizado_V25_Ref} 

\providecommand{\href}[2]{#2}\begingroup\raggedright\begin{thebibliography}{10}

\bibitem{Poschl:1933zz}
G.~P{\"o}schl and E.~Teller, ``{Bemerkungen zur Quantenmechanik des
  anharmonischen Oszillators},''
  \href{http://dx.doi.org/10.1007/BF01331132}{{\em Z. Phys.} {\bf 83} (1933)
  143--151}.

\bibitem{Ferrari:1984zz}
V.~Ferrari and B.~Mashhoon, ``{New approach to the quasinormal modes of a black
  hole},'' \href{http://dx.doi.org/10.1103/PhysRevD.30.295}{{\em Phys. Rev. D}
  {\bf 30} (1984)  295--304}.

\bibitem{Beyer:1998nu}
H.~R. Beyer, ``{On the completeness of the quasinormal modes of the
  Poschl-Teller potential},''
  \href{http://dx.doi.org/10.1007/s002200050651}{{\em Commun. Math. Phys.} {\bf
  204} (1999)  397--423}, \href{http://arxiv.org/abs/gr-qc/9803034}{{\tt
  arXiv:gr-qc/9803034}}.

\bibitem{QNM_dS}
D.-P. Du, B.~Wang, and R.-K. Su, ``{Quasinormal modes in pure de Sitter
  space-times},'' \href{http://dx.doi.org/10.1103/PhysRevD.70.064024}{{\em
  Phys. Rev. D} {\bf 70} (2004)  064024},
  \href{http://arxiv.org/abs/hep-th/0404047}{{\tt arXiv:hep-th/0404047}}.

\bibitem{Lopez-Ortega:2006aal}
A.~Lopez-Ortega, ``{Quasinormal modes of D-dimensional de Sitter spacetime},''
  \href{http://dx.doi.org/10.1007/s10714-006-0335-9}{{\em Gen. Rel. Grav.} {\bf
  38} (2006)  1565--1591}, \href{http://arxiv.org/abs/gr-qc/0605027}{{\tt
  arXiv:gr-qc/0605027}}.

\bibitem{cardona_2017_PT}
A.~F. Cardona and C.~Molina, ``{Quasinormal modes of generalized
  P\"oschl\textendash{}Teller potentials},''
  \href{http://dx.doi.org/10.1088/1361-6382/aa9428}{{\em Class. Quant. Grav.}
  {\bf 34} (2017) no.~24, 245002}, \href{http://arxiv.org/abs/1711.00479}{{\tt
  arXiv:1711.00479 [gr-qc]}}.

\bibitem{QNM_BH_Julio}
J.~C. Fabris, M.~G. Richarte, and A.~Saa, ``{Quasinormal modes and self-adjoint
  extensions of the Schr\"odinger operator},''
  \href{http://dx.doi.org/10.1103/PhysRevD.103.045001}{{\em Phys. Rev. D} {\bf
  103} (2021) no.~4, 045001}, \href{http://arxiv.org/abs/2010.10674}{{\tt
  arXiv:2010.10674 [gr-qc]}}.

\bibitem{deAlfaro:1976vlx}
V.~de~Alfaro, S.~Fubini, and G.~Furlan, ``{Conformal Invariance in Quantum
  Mechanics},'' \href{http://dx.doi.org/10.1007/BF02785666}{{\em Nuovo Cim. A}
  {\bf 34} (1976)  569}.

\bibitem{Calogero:1970nt}
F.~Calogero, ``{Solution of the one-dimensional N body problems with quadratic
  and/or inversely quadratic pair potentials},''
  \href{http://dx.doi.org/10.1063/1.1665604}{{\em J. Math. Phys.} {\bf 12}
  (1971)  419--436}.

\bibitem{Renor_Inver_Squa}
H.~E. Camblong, L.~N. Epele, H.~Fanchiotti, and C.~A. Garcia~Canal,
  ``{Renormalization of the inverse square potential},''
  \href{http://dx.doi.org/10.1103/PhysRevLett.85.1590}{{\em Phys. Rev. Lett.}
  {\bf 85} (2000)  1590--1593}, \href{http://arxiv.org/abs/hep-th/0003014}{{\tt
  arXiv:hep-th/0003014}}.

\bibitem{Camblong:2003mb}
H.~E. Camblong and C.~R. Ordonez, ``{Renormalization in conformal quantum
  mechanics},'' \href{http://dx.doi.org/10.1016/j.physleta.2005.06.110}{{\em
  Phys. Lett. A} {\bf 345} (2005)  22--30},
  \href{http://arxiv.org/abs/hep-th/0305035}{{\tt arXiv:hep-th/0305035}}.

\bibitem{Kaplan:2009kr}
D.~B. Kaplan, J.-W. Lee, D.~T. Son, and M.~A. Stephanov, ``{Conformality
  Lost},'' \href{http://dx.doi.org/10.1103/PhysRevD.80.125005}{{\em Phys. Rev.
  D} {\bf 80} (2009)  125005}, \href{http://arxiv.org/abs/0905.4752}{{\tt
  arXiv:0905.4752 [hep-th]}}.

\bibitem{renor_orig}
S.~R. Beane, P.~F. Bedaque, L.~Childress, A.~Kryjevski, J.~McGuire, and U.~van
  Kolck, ``{Singular potentials and limit cycles},''
  \href{http://dx.doi.org/10.1103/PhysRevA.64.042103}{{\em Phys. Rev. A} {\bf
  64} (2001)  042103}, \href{http://arxiv.org/abs/quant-ph/0010073}{{\tt
  arXiv:quant-ph/0010073}}.

\bibitem{renor_Group_Limit_Cycle}
E.~Braaten and D.~Phillips, ``{The Renormalization group limit cycle for the
  $1/r^2$ potential},''
  \href{http://dx.doi.org/10.1103/PhysRevA.70.052111}{{\em Phys. Rev. A} {\bf
  70} (2004)  052111}, \href{http://arxiv.org/abs/hep-th/0403168}{{\tt
  arXiv:hep-th/0403168}}.

\bibitem{renor_m_w_c}
D.~Bouaziz and M.~Bawin, ``Singular inverse-square potential: renormalization
  and self-adjoint extensions for medium to weak coupling,'' {\em Physical
  Review A} {\bf 89} (2014) no.~2, 022113.

\bibitem{Gitman}
D.~M. Gitman, I.~V. Tyutin, and B.~L. Voronov, ``{Self-adjoint extensions and
  spectral analysis in Calogero problem},''
  \href{http://arxiv.org/abs/0903.5277}{{\tt arXiv:0903.5277 [quant-ph]}}.

\bibitem{Ishibashi:2004wx}
A.~Ishibashi and R.~M. Wald, ``{Dynamics in nonglobally hyperbolic static
  space-times. 3. Anti-de Sitter space-time},''
  \href{http://dx.doi.org/10.1088/0264-9381/21/12/012}{{\em Class. Quant.
  Grav.} {\bf 21} (2004)  2981--3014},
  \href{http://arxiv.org/abs/hep-th/0402184}{{\tt arXiv:hep-th/0402184}}.

\bibitem{Efimov}
V.~N. Efimov, ``Weakly bound states of three resonantly interacting
  particles.,'' tech. rep., Ioffe Inst. of Physics and Tech., Leningrad, 1970.

\bibitem{Infeld:1951mw}
L.~Infeld and T.~E. Hull, ``{The factorization method},''
  \href{http://dx.doi.org/10.1103/RevModPhys.23.21}{{\em Rev. Mod. Phys.} {\bf
  23} (1951)  21--68}.

\bibitem{Cooper:1994eh}
F.~Cooper, A.~Khare, and U.~Sukhatme, ``{Supersymmetry and quantum
  mechanics},'' \href{http://dx.doi.org/10.1016/0370-1573(94)00080-M}{{\em
  Phys. Rept.} {\bf 251} (1995)  267--385},
  \href{http://arxiv.org/abs/hep-th/9405029}{{\tt arXiv:hep-th/9405029}}.

\bibitem{dutt1988}
R.~Dutt, A.~Khare, and U.~P. Sukhatme, ``{Supersymmetry, Shape Invariance and
  Exactly Solvable Potentials},'' \href{http://dx.doi.org/10.1119/1.15697}{{\em
  Am. J. Phys.} {\bf 56} (1988)  163--168}.

\bibitem{U_AB}
U.~Camara~da Silva, ``{Renormalization group flow of the Aharonov-Bohm
  scattering amplitude},''
  \href{http://dx.doi.org/10.1016/j.aop.2018.09.001}{{\em Annals Phys.} {\bf
  398} (2018)  38--54}.

\bibitem{Lima:2019xzg}
J.~V.~S. Scursulim, U.~Camara~da Silva, G.~M. Sotkov, and A.~A. Lima,
  ``{Supersymmetry shielding the scaling symmetry of conformal quantum
  mechanics},'' \href{http://dx.doi.org/10.1103/PhysRevA.101.032105}{{\em Phys.
  Rev. A} {\bf 101} (2020) no.~3, 032105},
  \href{http://arxiv.org/abs/1912.13014}{{\tt arXiv:1912.13014 [hep-th]}}.

\bibitem{Berezinsky:1972rfj}
V.~L. Berezinsky, ``{Destruction of Long-range Order in One-dimensional and
  Two-dimensional Systems Possessing a Continuous Symmetry Group. II. Quantum
  Systems.},'' {\em Sov. Phys. JETP} {\bf 34} (1972) no.~3, 610.

\bibitem{Berezinsky:1970fr}
V.~L. Berezinsky, ``{Destruction of Long-range Order in One-dimensional and
  Two-dimensional Systems Having a continuous symmetry group. I. Classical
  systems},'' {\em Sov. Phys. JETP} {\bf 32} (1971)  493--500.

\bibitem{NIST:DLMF}
``{\it NIST Digital Library of Mathematical Functions}.''
  \url{https://dlmf.nist.gov/}, release 1.1.10 of 2023-06-15.
\newblock \url{https://dlmf.nist.gov/}. F.~W.~J. Olver, A.~B. {Olde Daalhuis},
  D.~W. Lozier, B.~I. Schneider, R.~F. Boisvert, C.~W. Clark, B.~R. Miller,
  B.~V. Saunders, H.~S. Cohl, and M.~A. McClain, eds.

\bibitem{Zhidenko:2003wq}
A.~Zhidenko, ``{Quasinormal modes of Schwarzschild de Sitter black holes},''
  \href{http://dx.doi.org/10.1088/0264-9381/21/1/019}{{\em Class. Quant. Grav.}
  {\bf 21} (2004)  273--280}, \href{http://arxiv.org/abs/gr-qc/0307012}{{\tt
  arXiv:gr-qc/0307012}}.

\bibitem{Cardoso:2003sw}
V.~Cardoso and J.~P.~S. Lemos, ``{Quasinormal modes of the near extremal
  Schwarzschild-de Sitter black hole},''
  \href{http://dx.doi.org/10.1103/PhysRevD.67.084020}{{\em Phys. Rev. D} {\bf
  67} (2003)  084020}, \href{http://arxiv.org/abs/gr-qc/0301078}{{\tt
  arXiv:gr-qc/0301078}}.

\bibitem{Konoplya:2011qq}
R.~A. Konoplya and A.~Zhidenko, ``{Quasinormal modes of black holes: From
  astrophysics to string theory},''
  \href{http://dx.doi.org/10.1103/RevModPhys.83.793}{{\em Rev. Mod. Phys.} {\bf
  83} (2011)  793--836}, \href{http://arxiv.org/abs/1102.4014}{{\tt
  arXiv:1102.4014 [gr-qc]}}.

\bibitem{Berti:2009kk}
E.~Berti, V.~Cardoso, and A.~O. Starinets, ``{Quasinormal modes of black holes
  and black branes},''
  \href{http://dx.doi.org/10.1088/0264-9381/26/16/163001}{{\em Class. Quant.
  Grav.} {\bf 26} (2009)  163001}, \href{http://arxiv.org/abs/0905.2975}{{\tt
  arXiv:0905.2975 [gr-qc]}}.

\bibitem{Gango_book}
A.~Gangopadhyaya, J.~V. Mallow, and C.~Rasinariu,
  \href{http://dx.doi.org/10.1142/10475}{{\em {Supersymmetric Quantum
  Mechanics}: {An Introduction}}}.
\newblock World Scientific, 2017.

\bibitem{Odake2009}
S.~Odake and R.~Sasaki, ``{Infinitely many shape invariant potentials and new
  orthogonal polynomials},''
  \href{http://dx.doi.org/10.1016/j.physletb.2009.08.004}{{\em Phys. Lett. B}
  {\bf 679} (2009)  414--417}, \href{http://arxiv.org/abs/0906.0142}{{\tt
  arXiv:0906.0142 [math-ph]}}.

\bibitem{Odake2011}
S.~Odake and R.~Sasaki, ``{Exactly Solvable Quantum Mechanics and Infinite
  Families of Multi-indexed Orthogonal Polynomials},''
  \href{http://dx.doi.org/10.1016/j.physletb.2011.06.075}{{\em Phys. Lett. B}
  {\bf 702} (2011)  164--170}, \href{http://arxiv.org/abs/1105.0508}{{\tt
  arXiv:1105.0508 [math-ph]}}.

\bibitem{Odake2010}
S.~Odake and R.~Sasaki, ``{Infinitely many shape invariant potentials and cubic
  identities of the Laguerre and Jacobi polynomials},''
  \href{http://dx.doi.org/10.1063/1.3371248}{{\em J. Math. Phys.} {\bf 51}
  053513}, \href{http://arxiv.org/abs/0911.1585}{{\tt arXiv:0911.1585
  [math-ph]}}.

\bibitem{bougie2010}
J.~Bougie, A.~Gangopadhyaya, and J.~V. Mallow, ``{Generation of a Complete Set
  of Supersymmetric Shape Invariant Potentials from an Euler Equation},''
  \href{http://dx.doi.org/10.1103/PhysRevLett.105.210402}{{\em Phys. Rev.
  Lett.} {\bf 105} (2010)  210402}, \href{http://arxiv.org/abs/1008.2035}{{\tt
  arXiv:1008.2035 [hep-th]}}.

\bibitem{Chirenti:2012fr}
C.~Chirenti, A.~Saa, and J.~Skakala, ``{Quasinormal modes for the scattering on
  a naked Reissner-Nordstrom singularity},''
  \href{http://dx.doi.org/10.1103/PhysRevD.86.124008}{{\em Phys. Rev. D} {\bf
  86} (2012)  124008}, \href{http://arxiv.org/abs/1206.0037}{{\tt
  arXiv:1206.0037 [gr-qc]}}.

\bibitem{Robin_1}
T.~Harada, T.~Ishii, T.~Katagiri, and N.~Tanahashi, ``{Hairy black holes in AdS
  with Robin boundary conditions},''
  \href{http://arxiv.org/abs/2304.02267}{{\tt arXiv:2304.02267 [hep-th]}}.

\bibitem{Robin_2}
S.~Kinoshita, T.~Kozuka, K.~Murata, and K.~Sugawara, ``{Quasinormal mode
  spectrum of the AdS black hole with the Robin boundary condition},''
  \href{http://arxiv.org/abs/2305.17942}{{\tt arXiv:2305.17942 [gr-qc]}}.

\bibitem{Cooper:2001zd}
F.~Cooper, A.~Khare, and U.~Sukhatme,
  \href{http://dx.doi.org/https://doi.org/10.1142/4687}{{\em {Supersymmetry in
  Quantum Mechanics}}}.
\newblock World Scientific, 2001.

\bibitem{Molina:2003ff}
C.~Molina, ``{Quasinormal modes of d-dimensional spherical black holes with
  near extreme cosmological constant},''
  \href{http://dx.doi.org/10.1103/PhysRevD.68.064007}{{\em Phys. Rev. D} {\bf
  68} (2003)  064007}, \href{http://arxiv.org/abs/gr-qc/0304053}{{\tt
  arXiv:gr-qc/0304053}}.

\bibitem{Churilova:2021nnc}
M.~S. Churilova, R.~A. Konoplya, and A.~Zhidenko, ``{Analytic formula for
  quasinormal modes in the near-extreme Kerr-Newman\textendash{}de Sitter
  spacetime governed by a non-P\"oschl-Teller potential},''
  \href{http://dx.doi.org/10.1103/PhysRevD.105.084003}{{\em Phys. Rev. D} {\bf
  105} (2022) no.~8, 084003}, \href{http://arxiv.org/abs/2108.04858}{{\tt
  arXiv:2108.04858 [gr-qc]}}.

\bibitem{Jaramillo:2020tuu}
J.~L. Jaramillo, R.~Panosso~Macedo, and L.~Al~Sheikh, ``{Pseudospectrum and
  Black Hole Quasinormal Mode Instability},''
  \href{http://dx.doi.org/10.1103/PhysRevX.11.031003}{{\em Phys. Rev. X} {\bf
  11} (2021) no.~3, 031003}, \href{http://arxiv.org/abs/2004.06434}{{\tt
  arXiv:2004.06434 [gr-qc]}}.

\bibitem{mendoncca20152}
T.~S. Mendon{\c{c}}a and H.~P. de~Oliveira, ``A note about a new class of
  two-kinks,'' {\em Journal of High Energy Physics} {\bf 2015} (2015) no.~6,
  1--12.

\bibitem{mendoncca20151}
T.~S. Mendon{\c{c}}a and H.~P. de~Oliveira, ``{The collision of two-kinks
  defects},'' \href{http://dx.doi.org/10.1007/JHEP09(2015)120}{{\em JHEP} {\bf
  09} (2015)  120}, \href{http://arxiv.org/abs/1502.03870}{{\tt
  arXiv:1502.03870 [hep-th]}}.

\bibitem{mendoncca2019}
T.~S. Mendon{\c{c}}a and H.~D. Oliveira, ``The collision of two-kinks
  revisited: the creation of kinks and lump-like defects as metastable
  states,'' {\em Brazilian Journal of Physics} {\bf 49} (2019) no.~6, 914--922.

\bibitem{Zhong}
Y.~Zhong, ``{Singular P{\"o}schl$-$Teller II potentials and gravitating
  kinks},'' \href{http://dx.doi.org/10.1007/JHEP09(2022)165}{{\em JHEP} {\bf
  09} (2022)  165}, \href{http://arxiv.org/abs/2207.12681}{{\tt
  arXiv:2207.12681 [hep-th]}}.

\bibitem{Bazeia:2022yyv}
D.~Bazeia, J.~a. G.~F. Campos, and A.~Mohammadi, ``{Resonance mediated by
  fermions in kink-antikink collisions},''
  \href{http://dx.doi.org/10.1007/JHEP12(2022)085}{{\em JHEP} {\bf 12} (2022)
  085}, \href{http://arxiv.org/abs/2208.13261}{{\tt arXiv:2208.13261
  [hep-th]}}.

\end{thebibliography}\endgroup
\end{document}